\documentclass[prl,twocolumn,superscriptaddress,reprint,floatfix]{revtex4-1}
\pdfoutput=1
\usepackage{graphicx}
\usepackage{times}
\usepackage{physics}   
\usepackage{bm}        
\usepackage{amssymb}   
\usepackage{amsmath}
\usepackage{wasysym,scalerel}
\usepackage{subcaption}
\usepackage{xcolor}
\usepackage{float}
\usepackage{graphicx}
\usepackage{comment} 
\usepackage{caption,tabularx}
\usepackage{booktabs} 
\usepackage{hyperref}

\let\vec\mathbf
\begin{document}

\title{Entanglement Renyi negativity across a finite temperature transition: a  Monte Carlo study }

\author{Kai-Hsin Wu}
\email{kaihsinwu@gmail.com}
\affiliation{Department of Physics and Center of Theoretical Sciences, National Taiwan University, Taipei 10607, Taiwan}

\author{Tsung-Cheng Lu}
\email{tsl015@ucsd.edu}
\affiliation{Department of Physics, University of California at San Diego, La Jolla, CA 92093, USA}

\author{Chia-Min Chung}
\email{chiaminchung@gmail.com}
\affiliation{Department of Physics and Arnold Sommerfeld Center for Theoretical Physics, Ludwig-Maximilians-Universitat Munchen, Theresienstrasse 37, 80333 Munchen, Germany}

\author{Ying-Jer Kao}
\email{yjkao@phys.ntu.edu.tw}
\affiliation{Department of Physics and Center of Theoretical Sciences, National Taiwan University, Taipei 10607, Taiwan}
 
 \author{Tarun Grover}
 \email{tagrover@ucsd.edu}
 \affiliation{Department of Physics, University of California at San Diego, La Jolla, CA 92093, USA}
 
\date{\today}

\begin{abstract}
Quantum entanglement is fragile to thermal fluctuations, which raises the question whether finite temperature phase transitions support long-range entanglement similar to their zero temperature counterparts. Here we use quantum Monte Carlo simulations to  study  the third Renyi negativity, a generalization of entanglement negativity,  as a proxy of mixed-state entanglement in the 2D transverse field Ising model across its finite temperature phase transition. 
We find that the area-law coefficient of the Renyi negativity is singular across the transition, while its subleading constant is zero within the  statistical error. This  indicates that  the entanglement is short-ranged at the critical point despite a divergent correlation length. Renyi negativity in several exactly solvable models also shows qualitative similarities  to that in the 2D transverse field Ising model.
\end{abstract}

\maketitle

Long-range correlations in a quantum system can lead to long-range quantum entanglement. For example, the entanglement in the ground state of a 1+1-D conformal field theory (CFT) for a subregion of size $\ell$ takes the form $S \sim c \log \ell$  \cite{Callan94,Holzhey94,Calabrese04} , and thus is not expressible as sum of local terms close to the entangling boundary i.e. $S(2\ell) \neq S(\ell)$, underlining the long range nature of entanglement. Similarly,  the entanglement of a 2+1-D CFT for a circular bipartition of radius $R$ is given by $S \sim R - F$, where $F$ is a universal number that is not expressible in terms of correlation function of local operators, and again captures the long-range entanglement present in the ground state wavefunction \cite{myers2010, myers2011, casini2011, casini2012, casini2015, pufu2011a, pufu2011b}. At the same time, long-range correlations do not necessarily imply long-range entanglement as is evident by considering a classical Ising model at its finite temperature critical point - the entanglement is clearly zero in this system for any biparition despite the system being described by a (Euclidean) 2D Ising CFT.  A more interesting question is to  consider a quantum Hamiltonian in $d$ space dimensions at a finite temperature critical point. Now the system is described by the Gibbs state $\rho \propto e^{-\beta H}$, which is not a pure state. The critical exponents for this system are described by a $d$ dimensional classical field theory \cite{sachdev2011quantum} since the imaginary time direction is finite. What is the nature of quantum entanglement across such a transition? Does there exist any universal long-distance component of entanglement at this critical point? For simulating ground states of quantum states, the presence or absence of long-range entanglement has crucial implications for the computational resources required (see e.g.\cite{pollmann2009, Ulrich2011}). Therefore, answering these questions may have implications for the simultability of finite-temperature quantum systems with divergent correlation length. Although enormous progress has been made in last two decades in  understanding  entanglement of pure quantum states, very little is understood about the entanglement of interacting many-body quantum systems in mixed states such as the Gibbs state. In this paper, we will study a specific quantity called \textit{entanglement Renyi negativity} at a finite temperature critical point for a 2+1-D lattice model using quantum Monte Carlo (QMC) simulations, and make progress on some of these qualitative questions.

Given a density matrix $\rho$ on a bipartite Hilbert space $\mathcal{H}_A \otimes \mathcal{H}_B$, the two parties $A$ and $B$ are separable, i.e. unentangled, if and only if $\rho$ can be expressed as a convex combination of direct product states: $\rho=\sum_i P_i \rho^A_i\otimes \rho_i^B$. There exist several measures of entanglement that quantify how much a given state deviates from a separable state. Most of these measures require optimization over all possible states in the Hilbert space, making them intractable for many-body systems \cite{horodecki_revmodphys}. However, there does exist at least a mixed state entanglement measure called entanglement negativity\cite{vidal2002} (henceforth just ``negativity '' for brevity), which does not invoke any optimization. To define this quantity, consider a density matrix acting on the Hilbert space $\mathcal{H}_A \otimes \mathcal{H}_B$: $\rho=\sum_{A,B,A',B'}\rho_{A,B;A'B'}\ket{A}\ket{B}\bra{A'}\bra{B'}$, a partial transpose operation over $A$ gives $\rho^{T_A}=\sum_{A,B,A',B'}\rho_{A,B;A'B'}\ket{A'}\ket{B}\bra{A}\bra{B'}$. The negativity $E_N$ is then defined as $E_N=\log (\norm{\rho^{T_A}}_1)$. Although negativity can be zero for a entangled mixed state, a nonzero negativity necessarily implies the nonzero entanglement between the two parties.

In spite of being  computable without requiring any optimization, negativity is analytically tractable only in simple models such as free bosonic and fermionic systems \cite{audenaert2002entanglement, shapourian2017, Lu2019_gaussian, shapourian2019}, one-dimensional conformal field theories and integrable spin-chains \cite{calabrese2012, calabrese2015, wichterich2009, ruggiero2016}, and systems that have a tensor network representation such as commuting projector Hamiltonians \cite{vidal2013, castelnovo2013, castelnovo2018, Lu2018_singularity, gray2018fast}. It is thus desirable to devise a QMC scheme for large-scale simulation. However, the definition of negativity involves a matrix one norm, which impedes the construction of a QMC algorithm. Taking cue from a somewhat similar obstacle encountered in the evaluation of von Neumann entropy for pure states~\cite{kallin2010}, one approach to make progress is to instead define a Renyi version of negativity, dubbed \textit{Renyi negativity}, which involves the moment of the partial transposed density matrix. It was first introduced as an analytical tool to calculate negativity in the conformal field theory~\cite{calabrese2012,calabrese2013entanglement}, and was later implemented in a QMC simulation by the replica trick in Ref.~\cite{alba2013, Chiamin:2014repqmc} for a 1D spin-chain and the Bose-Hubbard model.

Here we present an extensive numerical study for Renyi negativity in the 2D transverse field Ising model (TFIM) using  QMC. In contrast to the 1D models in Refs.~\cite{alba2013, Chiamin:2014repqmc}, the 2D TFIM hosts a finite temperature  transition, which allows us to pose questions related to the universal form of mixed state entanglement across the transition.
Refs.~\cite{Lu2019_gaussian} studied the negativity in an exactly solvable model which is motivated by the mean-field description of TFIM. It was found that in this model, negativity takes the following form: $E_N = \alpha L - \gamma + O(1/L)$. Here $\alpha$, the area-law coefficient of negativity, is shown to be a singular function of the tuning parameter, and $\gamma$ is found to vanish exponentially with system size $\gamma \sim e^{-L/\xi_Q}$, where $\xi_Q$ defines a `quantum correlation length' that remains finite even at the transition, in strong contrast to  the physical correlation length $\xi$ that diverges at the critical point. Therefore, in this model, the long-distance part of negativity vanishes in the thermodynamical limit, even at the critical point. Taking cue from these results, we will employ QMC to study Renyi negativity in 2D TFIM, and study both the area-law coefficient as well as the long-distance, universal subleading term $\gamma$ for this model. On the technical front, we will introduce and implement an expanded ensemble QMC method to extract the subleading term which scales much more favorably than a direct implementation of conventional stochastic series expansion (SSE) approach of Ref.~\cite{alba2013, Chiamin:2014repqmc}.

\underline{Renyi negativity in simple models:} The Renyi negativity of index $n$ is defined as $ R_n= -\log \left(   \frac{  \tr{  \left(\rho^{T_A} \right)^n   }  }{ \tr\rho^n  }      \right)$. When $\rho$ is a pure state, $R_n$ is directly related to Renyi entanglement entropy $S_n$ by the relations: $R_n \propto S_n$ for odd $n$ and $R_n \propto S_{n/2}$ for even $n$. $R_n$ reduces to $-E_N$ with an analytic continuation by sending $n\to 1$ for even $n$ \cite{calabrese2012,calabrese2013entanglement}.

\begin{figure}[t]
	\includegraphics[width=\linewidth]{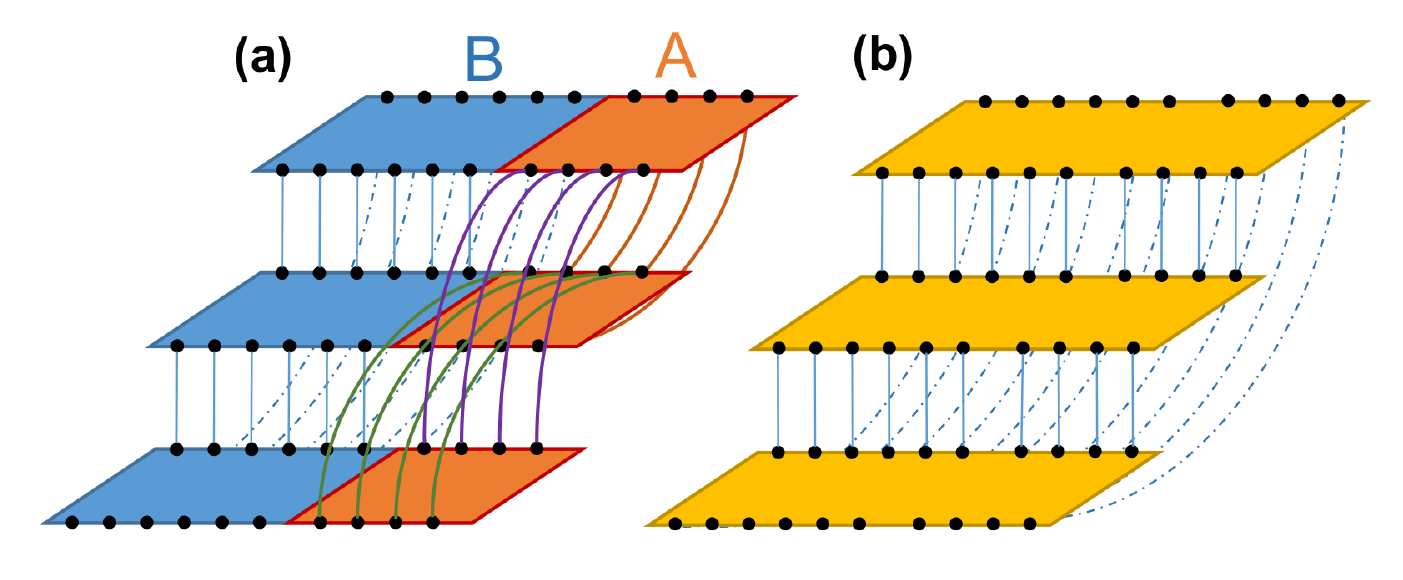}
	\caption{Boundary conditions for different replicas in  space-imaginary time for 	\textbf{(a)} $\tr{  \left(\rho^{T_A} \right)^3   }$ and	\textbf{(b)} $\tr (\rho^3)$.
			}
	\label{Fig:TopoZZ3} 
\end{figure}

For a large class of lattice models and field theories relevant to our discussion, Renyi negativity shares several key features with the negativity $E_N$.  For example, for the Gibbs state corresponding to a 1D conformal field theory, both $R_n$ and $E_N$ exhibit an area law with similar dependence on temperature: $E_N, R_n \sim \log(\beta)$ \cite{calabrese2012,calabrese2013entanglement}. Next, consider higher dimensional solvable models studied in Refs.~\cite{Lu2018_singularity, Lu2019_gaussian} that exhibit a  finite-temperature phase transition. The key results from these models were (i) For non-local models (such as the spherical model), $E_N$ is singular across the phase transition. (ii) For local models, area-law coefficient of $E_N$ is singular across the finite temperature  phase transition (iii) For local models, after subtracting off the local terms (which includes the area-law component), negativity decays exponentially even at the critical point: $\Delta E_N \sim e^{-L/\xi_Q}$ where $\xi_Q$ was called `quantum correlation length'. The significance of the last result is that it implies that the long-range component of negativity vanishes in the thermodynamic limit, in agreement with the conventional wisdom that these phase transitions are `classical' rather than `quantum'. 

We find all these features carry over to the Renyi negativity $R_n$, the main difference being that the temperature where the Renyi negativity $R_n$ is singular is given by $n T_c$ where $T_c$ is the actual critical temperature, i.e., the temperature where the partition function $Z = \tr \left(e^{- \beta H}\right)$ is singular. This is because $R_n$ involves raising the Gibbs state to the power $n$, and thus the effective inverse temperature for bulk of the system is given by $n\beta$, where $\beta$ is the physical inverse temperature. To illustrate these points, first consider the quantum spherical model from Ref.~\cite{Lu2018_singularity}, $H=\frac{1}{2} g \sum_{i=1}^N p_i^2 -\frac{1}{2N} \sum_{i,j=1}^N x_ix_j $, where $\{x_i\}$ is suject to the spherical constraint: $\delta(\frac{1}{N} \sum_{i=1}^N x_i^2-\frac{1}{4}  )$. This model hosts a finite-T transition at a coupling $g_c$ and temperature $T_c$ that satisfy the equation $2\sqrt{g_c} \coth(\frac{1}{2}\beta_c \sqrt{g_c}) = 1$.  We find that although the Renyi negativities for this model are continuous functions of temperature, the derivative $\frac{d R_n}{d T}$ is discontinuous at a temperature $n T_c$,  similar to the behavior of negativity $E_N$ \cite{supplement}. Since this model is non-local, Renyi negativities do not follow an area-law, and there is no distinction between local contributions to negativity from non-local ones. To that end, we next briefly report the results on Renyi negativity for a local model considered in Ref.~\cite{Lu2019_gaussian}:  $H=\frac{1}{2} \sum_{\vec{r}} \left( \pi_{\vec{r}}^2 +m^2 \phi_{\vec{r}}^2 \right) +  \frac{1}{2}  \sum_{\expval{\vec{r},\vec{r}'}}K \left( \phi_{\vec{r}} -\phi_{\vec{r}'}   \right) ^2,$ where the physical mass obeys $m = \sqrt{T-T_{n,c}}$ for $T>T_{n,c}$, and $m = \sqrt{2(T_{n,c}-T)}$ for $T<T_{n,c}$. Here $T_{n,c}=nT_c$ gives the critical temperature of the state $\rho\sim \exp{-n\beta H}$. This model can be considered a mean-field description of the TFIM while taking into account Gaussian fluctuations.  We find that the area-law coefficient of the Renyi negativity has a cusp singularity at a temperature $T = n T_c$ where $T_c$ is the physical critical temperature, while the subleading, long-distance part of Renyi negativity, defined via a subtraction scheme analogous to Kitaev-Preskill/Levin-Wen construction \cite{Kitaev06_1, levin2006detecting}, decays exponentially with system size, even at the critical point \cite{supplement}

\underline{Renyi negativity for 2+1-D transverse field Ising model:}
The models discussed above are exactly solvable, and one might wonder if the qualitative features exhibited by them may be attributed to this fact. We now turn our focus to the TFIM on a square lattice, which is known for hosting a finite temperature phase transition within 2D Ising universality class, and is not exactly solvable. The Hamiltonian is given by:
\begin{equation}
H=	-\sum_{\expval{ij}} \sigma^{z}_i \sigma^{z}_j - h_x \sum_{i} \sigma^{x}_i,
\end{equation}
where the $\sigma_i^z, \sigma_i^x$ are the Pauli-Z, Pauli-X operator at site $i$, and $\left< ij \right>$ denotes all the nearest neighbor pairs on a square lattice. We impose the periodic boundary condition, and set $h_x=2.75$. We first locate the corresponding critical inverse temperature $\beta_c = 1.0874(1)$ from a finite size scaling of the Binder ratio $B_2 =\left<M_z^4\right>/\left<M_z^2\right>^2$ calculated by the standard SSE  simulation \cite{supplement}. This result is consistent with previous QMC study\cite{Wessel:2016tfim}.

Since the Renyi negativity $R_n$ vanishes for $n=1, 2$, the smallest nontrivial integer is $n=3$, which will be the focus of our QMC simulations.  $R_3$ can be expressed as:
\begin{equation} \label{eq:Zratio}
R_3(A) = -\log \left(   \frac{  \tr{  \left(\rho^{T_A} \right)^3   }  }{ \tr\rho^3  } \right) =-\log \left(    \frac{Z[A,\beta,3]}{Z[3\beta]} \right),
\end{equation}
where $Z[A,\beta,3]= \tr \left\{    \left[   [\exp(-\beta H)]^{T_A}  \right]^3 \right\}     $ and $Z[3\beta] =\tr \left[\exp(-3\beta H) \right]  $ are the partition functions subjected to the boundary conditions shown in Figs.~\ref{Fig:TopoZZ3}(a) and (b) respectively. Therefore, the Renyi negativity can be calculated  using the SSE  by numerical integrating the difference between the energy estimators for different boundary conditions:
\begin{equation}\label{eq:R_3_integration}
R_3[\beta]=\int_{0}^{\beta}  d\beta' ~\expval{E(\beta')}_{A,\beta,3} -\expval{E(\beta')}_{3\beta},
\end{equation}
where $\expval{.}_{A,\beta,3}$ and $\expval{.}_{3\beta}$ denote the expectation values evaluated with corresponding boundary conditions.  Here, we focus on $dR_3/d\beta$ as the derivative enhances the singularity in a finite-size simulation. 
From Eq.~\eqref{eq:R_3_integration}, it is clear that $dR_3/d\beta$ corresponds simply to the difference between the energy estimators, therefore requiring no thermodynamic integration.

	\begin{figure}[t]
	\includegraphics[width=\linewidth]{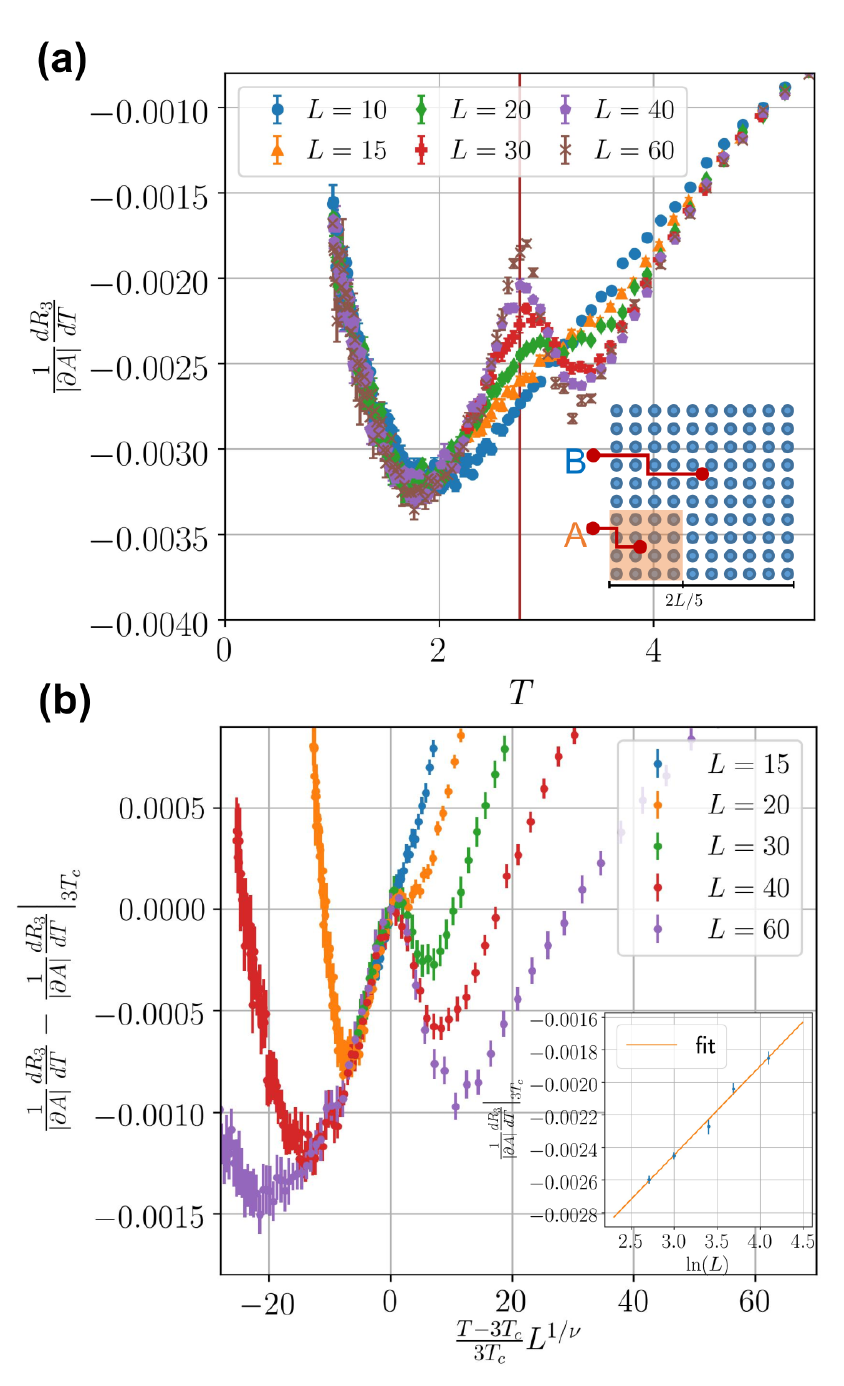}
	\caption{\textbf{(a)} Temperature derivative of the area law coefficient of the Renyi negativity across the finite-T transition. Geometry of the bipartition is shown in the inset and the vertical line indicates the location of the transition. \textbf{(b)} Data collapse for figure 2(a). The inset shows the linear scaling of temperature derivative at the critical point with $\log(L)$.} 
	\label{Fig:tR3dL} 
	\end{figure}

Fig.~\ref{Fig:tR3dL}(a) shows the temperature derivative of the area-law coefficient $R_3/|\partial A|$ as a function of the temperature for different system sizes. Here $|\partial A|$ denotes the length of the boundary of region $A$ over which partial transpose is taken.
The singularity occurs at $T = 3T_c$ consistent with our expectations. To understand the precise nature of this singularity, we note that on general symmetry grounds, the leading singular contribution to the area-law coefficient of negativity $E_N/|\partial A|$ as well as its Renyi counterparts such as $R_3$ will be proportional to the energy density \cite{Lu2019_gaussian}.  Therefore, $d (E_N/|\partial A|)/dT$ as well as $d (R_3/|\partial A|)/dT$ will receive a contribution proportional to the \textit{specific heat}. For instance, in the exactly solvable model discussed above, both $dE_N/dT$ and $dR_3/dT$ are discontinuous across the transition, which is indeed the singular behavior of the specific heat within mean-field \cite{supplement}. Returning to the 2D Ising model, we recall that the specific heat exponent $\alpha = 0$ and the correlation length exponent $\nu = 1$. Denoting linear size of the system by $L$ and $t = (T-T_c)/T_c$, the singular part of the specific heat in the vicinity of critical point takes the form $c_{v,\text{sing}}(L,t ) \sim c_{v,\text{sing}}(L,0) + f(Lt)$ where $c_{v,\text{sing}}(L,0) \propto \log(L)$  and $f$ is a universal function with the form $f(|x| \ll 1) \sim$ constant, and $f(|x| \gg 1) \sim  -\log(|x|)$ \cite{fisher1969}. Note that were $\alpha \neq 0$ (e.g. in the 3D Ising model), $c_{v,\text{sing}}(L,t )$ would take a different form, namely, $c_{v,\text{sing}}(L,t ) \sim c_{v,\text{sing}}(L,0)g(Lt)$.

Fig.2(b) shows the scaling collapse of $d(R_3/|\partial A|)/dT - d(R_3/|\partial A|)/dT\big|_{3T_c}$ with respect to $Lt$, where $ t =   (T-3T_c)/3T_c$, consistent with our expectation that $d(R_3/|\partial A|)/dT$ is proportional to the specific heat of the 2D Ising model. The inset shows the scaling right at the critical point, where we find that  $d(R_3/|\partial A|)/dT \propto \log(L)$, again consistent with 2D Ising universality.

\begin{figure}[t]
	\includegraphics[width=\linewidth]{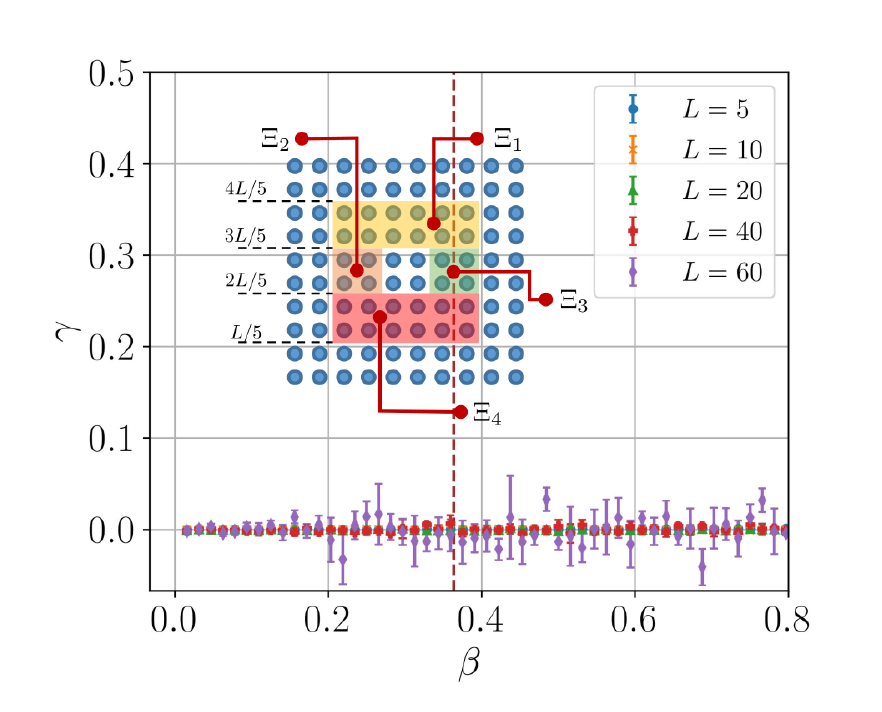}
	\caption{ The subleading contribution  $\gamma$ to the third  Renyi negativity $R_3$ obtained via Levin-Wen's subtraction scheme across the critical temperature. The inset shows the four sub-parts $\Xi_1$, $\Xi_2$, $\Xi_3$ and $\Xi_4$ employed in the subtraction scheme (see the main text for details). The dashed vertical line shows the location of the critical point.}
	\label{Fig:LWRes} 
\end{figure}

\underline{Universal long-range Renyi negativity:} So far we have demonstrated the Renyi negativity is singular across the finite temperature transition in 2D TFIM. 
Now we turn to the question whether there is a \textit{universal} subleading term in the Renyi negativity that reflects long-range quantum entanglement.
Writing $R_3 = a L - \gamma + b/L + \ldots$, where $L$ is the size of the entangling boundary, we are interested in whether $\gamma$ is non-zero. 
To extract $\gamma$ we use a subtraction scheme introduced by Levin and Wen in Ref.~\cite{levin2006detecting} in the context of ground state topological order, to cancel out the short-distance (local) contributions to negativity.
In particular, we construct four sub-regions $S_1$, $S_2$, $S_3$ and $S_4$ using combinations of four sub-parts marked as $\Xi_1$, $\Xi_2$, $\Xi_3$ and $\Xi_4$~(see inset of Fig.~\ref{Fig:LWRes}). 
The sub-regions $S_i$ are defined as $ S_1 \equiv \Xi_1 \cup \Xi_4, S_2 \equiv \Xi_1 \cup \Xi_2 \cup \Xi_4,   
  S_3 \equiv \Xi_1 \cup \Xi_2 \cup \Xi_3 \cup \Xi_4$, and $S_4 \equiv \Xi_1 \cup \Xi_3 \cup \Xi_4$. The non-local component $\gamma$ of $R_3$ is given by

\begin{align}
		\gamma &= -\left[ R_3(S_2)  - R_3(S_1) - R_3(S_3)+ R_3(S_4))\right]/2\nonumber\\
		&=-\left[ 2 R_3(S_2)  - R_3(S_1) - R_3(S_3)\right]/2,
		               \label{eq:red}
\end{align}
where we have used the relation $R_3\left(S_2\right)=R_3\left(S_4\right)$ arising from the symmetry of the model Hamiltonian.
	
The most straightforward way to compute $\gamma$ is to calculate $R_3(S_i)$ separately and perform the subtraction as in Eq.~\eqref{eq:red}. However, this requires three independent simulations and, the errors from each $R_3(S_i)$ will cumulate in the final subtraction.
Here we develop an expanded ensemble method that allows us to calculate $\gamma$ in a \emph{single} simulation. We first write $\gamma$ as the logarithm of the ratio of  partition functions
	\begin{equation}
		\gamma = \frac{1}{2} \log \frac{Z_{S_2}^2}{  Z_{S_1}   Z_{S_3} },
	\end{equation} 
where $Z_{S_i}$ is a shorthand notation for $Z[S_i,\beta,3]$.

To implement our method, in addition to the conventional SSE update, we also perform sampling in an expanded ensemble of the partition functions. In particular, we allow the system to switch between different partition functions $Z_{S_i}$ by changing the imaginary-time boundary conditions (see Fig.~\ref{Fig:TopoZZ3}(a)).
This can be achieved by sampling the total partition function $Z_\text{tot}$ defined as,
\begin{equation}
Z_\text{tot} = \sum_{i=1}^3 Z_{S_i},
\end{equation}
by proposing a move from $Z_{S_i}$ to either  $Z_{S_{i+1}}$  or $Z_{S_{i-1}}$ with equal probability. 
The update is  accepted if the spin configuration is consistent with the new boundary conditions.
It is clear that these moves correspond to adding or removing only region $\Xi_2$ or $\Xi_3$, which is much smaller than $S_i$, so a better acceptance rate can be achieved. 
The ratio $\frac{Z_{S_2}^2}{Z_{S_1}Z_{S_3}}$ then is simply estimated by $\frac{N_{S_2}^2}{N_{S_1}N_{S_3}}$, where $N_{S_i}$ is the number of samples in $Z_{S_i}$.

Since $\gamma$ is computed in a single simulation with an enlarged ensemble, we avoid the accumulation of error in the naive post-subtraction.
The new method is crucial in obtaining accurate $\gamma$, especially for the  large system size $L=60$.

	As the system size increases, the acceptance rates for exchanging regions $\Xi_2$ and $\Xi_3$ becomes smaller as more sites need to be updated. In such a case, we  can further divide $\Xi$ into several smaller subregions to add more intermediate ensembles and  optimize the performance with the re-weighting method \cite{supplement}. The simulation typically runs with $10^8$ Monte Carlo steps for smaller system sizes, and runs with around $10^9$ Monte Carlo steps for larger system sizes.
	
 Figure~\ref{Fig:LWRes} shows the results for $\gamma$. It is essentially zero at  temperature across the transition for all the system sizes  we consider, despite the fact that each individual term $R_3(S_i)$ that enters the Levin-Wen subtraction is singular at the transition (Fig.~\ref{Fig:tR3dL}).
This indicates that this finite-temperature transition is driven purely by  classical correlations and there exists no long-range entanglement at the transition, in line with our expectations based on the results from Ref.~\cite{Lu2019_gaussian} and of the exactly solvable models discussed above.

\underline{ \textit{Conclusion}:} We presented a first  QMC study of the Renyi negativity, a variant of negativity, across a finite-temperature phase transition in a non-integrable model, namely the two dimensional transverse field Ising model. We found a clear signature of singularity in the area-law coefficient of bipartite Renyi negativity, and perhaps more interestingly, vanishing of the subleading, non-local part of Renyi negativity. This indicates that the long-range correlations inherent to the critical point are completely classical, and the singularity associated with quantum correlations is localized close to the boundary. To extract this subleading term, we implemented the Levin-Wen subtraction scheme using a novel Monte Carlo algorithm that automatically cancels out the leading area-law contribution in a single simulation. 

We note that Ref.\cite{sherman2016} used a linked-cluster expansion to argue that the area-law coefficient of negativity is $\textit{not}$ singular across the finite-T transition in the 2D transverse field Ising model. Although we only studied  Renyi negativity, our results along with the results of Ref.\cite{Lu2019_gaussian}  strongly suggest that the lack of any visible singularity in Ref.\cite{sherman2016} is due to rather small system sizes accessible within the linked-cluster expansion  ($L \lesssim 10$). Even for the Renyi negativity, singularity at the critical point would not be visible at such sizes.

We also extended the analytical results on the negativity of exactly solvable results to the Renyi negativity, and found that they share essentially all qualitative features close to a finite temperature  transition. In particular, while the area-law coefficient is singular, the subleading component $\gamma$ vanishes exponentially with the system size: $\gamma \sim e^{-L/\xi_Q}$ where $\xi_Q \sim \beta$ is the quantum correlation length. We are unable to do similar scaling analysis for the 2D transverse field Ising model because  the Monte Carlo sampling error in $\gamma$ increases rapidly when increasing the system size while the mean value of $\gamma$ is close to zero. 

The vanishing of the non-local component of Renyi negativity suggests that the Gibbs state is separable up to short-distance quantum correlations. Therefore, we expect that there exists a `minimally entangled typical thermal state' (METTS) decomposition \cite{Stoudenmire_2010} of the Gibbs state both near and at the finite temperature  transition: $\rho = \sum_i p_i |\psi_i \rangle \langle \psi_i|$ where each pure state $|\psi_i\rangle$ is short-range entangled. Another promising future direction would be  to study the Renyi negativity in 4D toric code, which is argued to host a finite temperature transition from a topological ordered phase to a topologically trivial Gibbs state \cite{dennis2002}, using a similar QMC scheme. Finally, it will be interesting to apply the expanded ensemble method to extract topological entanglement entropy in gapped ground states.

\section{Acknowledgment}
We thank Dan Arovas and John McGreevy for helpful discussions. This work was supported by Ministry of Science and Technology (MOST) of Taiwan under GrantsNew numbers  108-2112-M-002 -020 -MY3, 107-2112-M-002 -016 -MY3, and partly supported by National Center of Theoretical Science (NCTS) of Taiwan with Young Scholar-fellowship to K.H.W. We are grateful to the National Center for High-performance Computing for computer time and facilities. 	TG is supported by an Alfred P. Sloan Research Fellowship,  National Science Foundation under Grant No. DMR-1752417, and the University of California's Multicampus Research Programs and Initiatives  (MRP-19-601445).

\bibliographystyle{apsrev4-1}

\bibliography{refENTFIM}

\begin{thebibliography}{42}%
\makeatletter
\providecommand \@ifxundefined [1]{%
 \@ifx{#1\undefined}
}%
\providecommand \@ifnum [1]{%
 \ifnum #1\expandafter \@firstoftwo
 \else \expandafter \@secondoftwo
 \fi
}%
\providecommand \@ifx [1]{%
 \ifx #1\expandafter \@firstoftwo
 \else \expandafter \@secondoftwo
 \fi
}%
\providecommand \natexlab [1]{#1}%
\providecommand \enquote  [1]{``#1''}%
\providecommand \bibnamefont  [1]{#1}%
\providecommand \bibfnamefont [1]{#1}%
\providecommand \citenamefont [1]{#1}%
\providecommand \href@noop [0]{\@secondoftwo}%
\providecommand \href [0]{\begingroup \@sanitize@url \@href}%
\providecommand \@href[1]{\@@startlink{#1}\@@href}%
\providecommand \@@href[1]{\endgroup#1\@@endlink}%
\providecommand \@sanitize@url [0]{\catcode `\\12\catcode `\$12\catcode
  `\&12\catcode `\#12\catcode `\^12\catcode `\_12\catcode `\%12\relax}%
\providecommand \@@startlink[1]{}%
\providecommand \@@endlink[0]{}%
\providecommand \url  [0]{\begingroup\@sanitize@url \@url }%
\providecommand \@url [1]{\endgroup\@href {#1}{\urlprefix }}%
\providecommand \urlprefix  [0]{URL }%
\providecommand \Eprint [0]{\href }%
\providecommand \doibase [0]{http://dx.doi.org/}%
\providecommand \selectlanguage [0]{\@gobble}%
\providecommand \bibinfo  [0]{\@secondoftwo}%
\providecommand \bibfield  [0]{\@secondoftwo}%
\providecommand \translation [1]{[#1]}%
\providecommand \BibitemOpen [0]{}%
\providecommand \bibitemStop [0]{}%
\providecommand \bibitemNoStop [0]{.\EOS\space}%
\providecommand \EOS [0]{\spacefactor3000\relax}%
\providecommand \BibitemShut  [1]{\csname bibitem#1\endcsname}%
\let\auto@bib@innerbib\@empty
\bibitem [{\citenamefont {Callan}\ and\ \citenamefont
  {Wilczek}(1994)}]{Callan94}%
  \BibitemOpen
  \bibfield  {author} {\bibinfo {author} {\bibfnamefont {C.}~\bibnamefont
  {Callan}}\ and\ \bibinfo {author} {\bibfnamefont {F.}~\bibnamefont
  {Wilczek}},\ }\href {\doibase http://dx.doi.org/10.1016/0370-2693(94)91007-3}
  {\bibfield  {journal} {\bibinfo  {journal} {Physics Letters B}\ }\textbf
  {\bibinfo {volume} {333}},\ \bibinfo {pages} {55 } (\bibinfo {year}
  {1994})}\BibitemShut {NoStop}%
\bibitem [{\citenamefont {Holzhey}\ \emph {et~al.}(1994)\citenamefont
  {Holzhey}, \citenamefont {Larsen},\ and\ \citenamefont
  {Wilczek}}]{Holzhey94}%
  \BibitemOpen
  \bibfield  {author} {\bibinfo {author} {\bibfnamefont {C.}~\bibnamefont
  {Holzhey}}, \bibinfo {author} {\bibfnamefont {F.}~\bibnamefont {Larsen}}, \
  and\ \bibinfo {author} {\bibfnamefont {F.}~\bibnamefont {Wilczek}},\ }\href
  {\doibase http://dx.doi.org/10.1016/0550-3213(94)90402-2} {\bibfield
  {journal} {\bibinfo  {journal} {Nuclear Physics B}\ }\textbf {\bibinfo
  {volume} {424}},\ \bibinfo {pages} {443 } (\bibinfo {year}
  {1994})}\BibitemShut {NoStop}%
\bibitem [{\citenamefont {Calabrese}\ and\ \citenamefont
  {Cardy}(2004)}]{Calabrese04}%
  \BibitemOpen
  \bibfield  {author} {\bibinfo {author} {\bibfnamefont {P.}~\bibnamefont
  {Calabrese}}\ and\ \bibinfo {author} {\bibfnamefont {J.}~\bibnamefont
  {Cardy}},\ }\href {http://stacks.iop.org/1742-5468/2004/i=06/a=P06002}
  {\bibfield  {journal} {\bibinfo  {journal} {Journal of Statistical Mechanics:
  Theory and Experiment}\ }\textbf {\bibinfo {volume} {2004}},\ \bibinfo
  {pages} {P06002} (\bibinfo {year} {2004})}\BibitemShut {NoStop}%
\bibitem [{\citenamefont {{Myers}}\ and\ \citenamefont
  {{Sinha}}(2010)}]{myers2010}%
  \BibitemOpen
  \bibfield  {author} {\bibinfo {author} {\bibfnamefont {R.~C.}\ \bibnamefont
  {{Myers}}}\ and\ \bibinfo {author} {\bibfnamefont {A.}~\bibnamefont
  {{Sinha}}},\ }\href {\doibase 10.1103/PhysRevD.82.046006} {\bibfield
  {journal} {\bibinfo  {journal} {\prd}\ }\textbf {\bibinfo {volume} {82}},\
  \bibinfo {eid} {046006} (\bibinfo {year} {2010})},\ \Eprint
  {http://arxiv.org/abs/1006.1263} {arXiv:1006.1263 [hep-th]} \BibitemShut
  {NoStop}%
\bibitem [{\citenamefont {{Myers}}\ and\ \citenamefont
  {{Sinha}}(2011)}]{myers2011}%
  \BibitemOpen
  \bibfield  {author} {\bibinfo {author} {\bibfnamefont {R.~C.}\ \bibnamefont
  {{Myers}}}\ and\ \bibinfo {author} {\bibfnamefont {A.}~\bibnamefont
  {{Sinha}}},\ }\href {\doibase 10.1007/JHEP01(2011)125} {\bibfield  {journal}
  {\bibinfo  {journal} {Journal of High Energy Physics}\ }\textbf {\bibinfo
  {volume} {2011}},\ \bibinfo {eid} {125} (\bibinfo {year} {2011})},\ \Eprint
  {http://arxiv.org/abs/1011.5819} {arXiv:1011.5819 [hep-th]} \BibitemShut
  {NoStop}%
\bibitem [{\citenamefont {{Casini}}\ \emph {et~al.}(2011)\citenamefont
  {{Casini}}, \citenamefont {{Huerta}},\ and\ \citenamefont
  {{Myers}}}]{casini2011}%
  \BibitemOpen
  \bibfield  {author} {\bibinfo {author} {\bibfnamefont {H.}~\bibnamefont
  {{Casini}}}, \bibinfo {author} {\bibfnamefont {M.}~\bibnamefont {{Huerta}}},
  \ and\ \bibinfo {author} {\bibfnamefont {R.~C.}\ \bibnamefont {{Myers}}},\
  }\href {\doibase 10.1007/JHEP05(2011)036} {\bibfield  {journal} {\bibinfo
  {journal} {Journal of High Energy Physics}\ }\textbf {\bibinfo {volume}
  {2011}},\ \bibinfo {eid} {36} (\bibinfo {year} {2011})},\ \Eprint
  {http://arxiv.org/abs/1102.0440} {arXiv:1102.0440 [hep-th]} \BibitemShut
  {NoStop}%
\bibitem [{\citenamefont {{Casini}}\ and\ \citenamefont
  {{Huerta}}(2012)}]{casini2012}%
  \BibitemOpen
  \bibfield  {author} {\bibinfo {author} {\bibfnamefont {H.}~\bibnamefont
  {{Casini}}}\ and\ \bibinfo {author} {\bibfnamefont {M.}~\bibnamefont
  {{Huerta}}},\ }\href {\doibase 10.1103/PhysRevD.85.125016} {\bibfield
  {journal} {\bibinfo  {journal} {\prd}\ }\textbf {\bibinfo {volume} {85}},\
  \bibinfo {eid} {125016} (\bibinfo {year} {2012})},\ \Eprint
  {http://arxiv.org/abs/1202.5650} {arXiv:1202.5650 [hep-th]} \BibitemShut
  {NoStop}%
\bibitem [{\citenamefont {{Casini}}\ \emph {et~al.}(2015)\citenamefont
  {{Casini}}, \citenamefont {{Huerta}}, \citenamefont {{Myers}},\ and\
  \citenamefont {{Yale}}}]{casini2015}%
  \BibitemOpen
  \bibfield  {author} {\bibinfo {author} {\bibfnamefont {H.}~\bibnamefont
  {{Casini}}}, \bibinfo {author} {\bibfnamefont {M.}~\bibnamefont {{Huerta}}},
  \bibinfo {author} {\bibfnamefont {R.~C.}\ \bibnamefont {{Myers}}}, \ and\
  \bibinfo {author} {\bibfnamefont {A.}~\bibnamefont {{Yale}}},\ }\href
  {\doibase 10.1007/JHEP10(2015)003} {\bibfield  {journal} {\bibinfo  {journal}
  {Journal of High Energy Physics}\ }\textbf {\bibinfo {volume} {2015}},\
  \bibinfo {eid} {3} (\bibinfo {year} {2015})},\ \Eprint
  {http://arxiv.org/abs/1506.06195} {arXiv:1506.06195 [hep-th]} \BibitemShut
  {NoStop}%
\bibitem [{\citenamefont {{Jafferis}}\ \emph {et~al.}(2011)\citenamefont
  {{Jafferis}}, \citenamefont {{Klebanov}}, \citenamefont {{Pufu}},\ and\
  \citenamefont {{Safdi}}}]{pufu2011a}%
  \BibitemOpen
  \bibfield  {author} {\bibinfo {author} {\bibfnamefont {D.~L.}\ \bibnamefont
  {{Jafferis}}}, \bibinfo {author} {\bibfnamefont {I.~R.}\ \bibnamefont
  {{Klebanov}}}, \bibinfo {author} {\bibfnamefont {S.~S.}\ \bibnamefont
  {{Pufu}}}, \ and\ \bibinfo {author} {\bibfnamefont {B.~R.}\ \bibnamefont
  {{Safdi}}},\ }\href {\doibase 10.1007/JHEP06(2011)102} {\bibfield  {journal}
  {\bibinfo  {journal} {Journal of High Energy Physics}\ }\textbf {\bibinfo
  {volume} {2011}},\ \bibinfo {eid} {102} (\bibinfo {year} {2011})},\ \Eprint
  {http://arxiv.org/abs/1103.1181} {arXiv:1103.1181 [hep-th]} \BibitemShut
  {NoStop}%
\bibitem [{\citenamefont {{Klebanov}}\ \emph {et~al.}(2011)\citenamefont
  {{Klebanov}}, \citenamefont {{Pufu}},\ and\ \citenamefont
  {{Safdi}}}]{pufu2011b}%
  \BibitemOpen
  \bibfield  {author} {\bibinfo {author} {\bibfnamefont {I.~R.}\ \bibnamefont
  {{Klebanov}}}, \bibinfo {author} {\bibfnamefont {S.~S.}\ \bibnamefont
  {{Pufu}}}, \ and\ \bibinfo {author} {\bibfnamefont {B.~R.}\ \bibnamefont
  {{Safdi}}},\ }\href {\doibase 10.1007/JHEP10(2011)038} {\bibfield  {journal}
  {\bibinfo  {journal} {Journal of High Energy Physics}\ }\textbf {\bibinfo
  {volume} {2011}},\ \bibinfo {eid} {38} (\bibinfo {year} {2011})},\ \Eprint
  {http://arxiv.org/abs/1105.4598} {arXiv:1105.4598 [hep-th]} \BibitemShut
  {NoStop}%
\bibitem [{\citenamefont {Sachdev}(2011)}]{sachdev2011quantum}%
  \BibitemOpen
  \bibfield  {author} {\bibinfo {author} {\bibfnamefont {S.}~\bibnamefont
  {Sachdev}},\ }\href@noop {} {\emph {\bibinfo {title} {Quantum phase
  transitions}}}\ (\bibinfo  {publisher} {Cambridge university press},\
  \bibinfo {year} {2011})\BibitemShut {NoStop}%
\bibitem [{\citenamefont {Pollmann}\ \emph {et~al.}(2009)\citenamefont
  {Pollmann}, \citenamefont {Mukerjee}, \citenamefont {Turner},\ and\
  \citenamefont {Moore}}]{pollmann2009}%
  \BibitemOpen
  \bibfield  {author} {\bibinfo {author} {\bibfnamefont {F.}~\bibnamefont
  {Pollmann}}, \bibinfo {author} {\bibfnamefont {S.}~\bibnamefont {Mukerjee}},
  \bibinfo {author} {\bibfnamefont {A.~M.}\ \bibnamefont {Turner}}, \ and\
  \bibinfo {author} {\bibfnamefont {J.~E.}\ \bibnamefont {Moore}},\ }\href
  {\doibase 10.1103/PhysRevLett.102.255701} {\bibfield  {journal} {\bibinfo
  {journal} {Phys. Rev. Lett.}\ }\textbf {\bibinfo {volume} {102}},\ \bibinfo
  {pages} {255701} (\bibinfo {year} {2009})}\BibitemShut {NoStop}%
\bibitem [{\citenamefont {{Schollw{\"o}ck}}(2011)}]{Ulrich2011}%
  \BibitemOpen
  \bibfield  {author} {\bibinfo {author} {\bibfnamefont {U.}~\bibnamefont
  {{Schollw{\"o}ck}}},\ }\href {\doibase 10.1016/j.aop.2010.09.012} {\bibfield
  {journal} {\bibinfo  {journal} {Annals of Physics}\ }\textbf {\bibinfo
  {volume} {326}},\ \bibinfo {pages} {96} (\bibinfo {year} {2011})},\ \Eprint
  {http://arxiv.org/abs/1008.3477} {arXiv:1008.3477 [cond-mat.str-el]}
  \BibitemShut {NoStop}%
\bibitem [{\citenamefont {Horodecki}\ \emph {et~al.}(2009)\citenamefont
  {Horodecki}, \citenamefont {Horodecki}, \citenamefont {Horodecki},\ and\
  \citenamefont {Horodecki}}]{horodecki_revmodphys}%
  \BibitemOpen
  \bibfield  {author} {\bibinfo {author} {\bibfnamefont {R.}~\bibnamefont
  {Horodecki}}, \bibinfo {author} {\bibfnamefont {P.}~\bibnamefont
  {Horodecki}}, \bibinfo {author} {\bibfnamefont {M.}~\bibnamefont
  {Horodecki}}, \ and\ \bibinfo {author} {\bibfnamefont {K.}~\bibnamefont
  {Horodecki}},\ }\href {\doibase 10.1103/RevModPhys.81.865} {\bibfield
  {journal} {\bibinfo  {journal} {Rev. Mod. Phys.}\ }\textbf {\bibinfo {volume}
  {81}},\ \bibinfo {pages} {865} (\bibinfo {year} {2009})}\BibitemShut
  {NoStop}%
\bibitem [{\citenamefont {Vidal}\ and\ \citenamefont
  {Werner}(2002)}]{vidal2002}%
  \BibitemOpen
  \bibfield  {author} {\bibinfo {author} {\bibfnamefont {G.}~\bibnamefont
  {Vidal}}\ and\ \bibinfo {author} {\bibfnamefont {R.~F.}\ \bibnamefont
  {Werner}},\ }\href {\doibase 10.1103/PhysRevA.65.032314} {\bibfield
  {journal} {\bibinfo  {journal} {Phys. Rev. A}\ }\textbf {\bibinfo {volume}
  {65}},\ \bibinfo {pages} {032314} (\bibinfo {year} {2002})}\BibitemShut
  {NoStop}%
\bibitem [{\citenamefont {Audenaert}\ \emph {et~al.}(2002)\citenamefont
  {Audenaert}, \citenamefont {Eisert}, \citenamefont {Plenio},\ and\
  \citenamefont {Werner}}]{audenaert2002entanglement}%
  \BibitemOpen
  \bibfield  {author} {\bibinfo {author} {\bibfnamefont {K.}~\bibnamefont
  {Audenaert}}, \bibinfo {author} {\bibfnamefont {J.}~\bibnamefont {Eisert}},
  \bibinfo {author} {\bibfnamefont {M.}~\bibnamefont {Plenio}}, \ and\ \bibinfo
  {author} {\bibfnamefont {R.}~\bibnamefont {Werner}},\ }\href@noop {}
  {\bibfield  {journal} {\bibinfo  {journal} {Physical Review A}\ }\textbf
  {\bibinfo {volume} {66}},\ \bibinfo {pages} {042327} (\bibinfo {year}
  {2002})}\BibitemShut {NoStop}%
\bibitem [{\citenamefont {Shapourian}\ \emph {et~al.}(2017)\citenamefont
  {Shapourian}, \citenamefont {Shiozaki},\ and\ \citenamefont
  {Ryu}}]{shapourian2017}%
  \BibitemOpen
  \bibfield  {author} {\bibinfo {author} {\bibfnamefont {H.}~\bibnamefont
  {Shapourian}}, \bibinfo {author} {\bibfnamefont {K.}~\bibnamefont
  {Shiozaki}}, \ and\ \bibinfo {author} {\bibfnamefont {S.}~\bibnamefont
  {Ryu}},\ }\href {\doibase 10.1103/PhysRevB.95.165101} {\bibfield  {journal}
  {\bibinfo  {journal} {Phys. Rev. B}\ }\textbf {\bibinfo {volume} {95}},\
  \bibinfo {pages} {165101} (\bibinfo {year} {2017})}\BibitemShut {NoStop}%
\bibitem [{\citenamefont {{Lu}}\ and\ \citenamefont
  {{Grover}}(2019)}]{Lu2019_gaussian}%
  \BibitemOpen
  \bibfield  {author} {\bibinfo {author} {\bibfnamefont {T.-C.}\ \bibnamefont
  {{Lu}}}\ and\ \bibinfo {author} {\bibfnamefont {T.}~\bibnamefont
  {{Grover}}},\ }\href@noop {} {\bibfield  {journal} {\bibinfo  {journal}
  {arXiv e-prints}\ ,\ \bibinfo {eid} {arXiv:1907.01569}} (\bibinfo {year}
  {2019})},\ \Eprint {http://arxiv.org/abs/1907.01569} {arXiv:1907.01569
  [cond-mat.str-el]} \BibitemShut {NoStop}%
\bibitem [{\citenamefont {Shapourian}\ and\ \citenamefont
  {Ryu}(2019)}]{shapourian2019}%
  \BibitemOpen
  \bibfield  {author} {\bibinfo {author} {\bibfnamefont {H.}~\bibnamefont
  {Shapourian}}\ and\ \bibinfo {author} {\bibfnamefont {S.}~\bibnamefont
  {Ryu}},\ }\href {\doibase 10.1088/1742-5468/ab11e0} {\bibfield  {journal}
  {\bibinfo  {journal} {Journal of Statistical Mechanics: Theory and
  Experiment}\ }\textbf {\bibinfo {volume} {2019}},\ \bibinfo {pages} {043106}
  (\bibinfo {year} {2019})}\BibitemShut {NoStop}%
\bibitem [{\citenamefont {Calabrese}\ \emph {et~al.}(2012)\citenamefont
  {Calabrese}, \citenamefont {Cardy},\ and\ \citenamefont
  {Tonni}}]{calabrese2012}%
  \BibitemOpen
  \bibfield  {author} {\bibinfo {author} {\bibfnamefont {P.}~\bibnamefont
  {Calabrese}}, \bibinfo {author} {\bibfnamefont {J.}~\bibnamefont {Cardy}}, \
  and\ \bibinfo {author} {\bibfnamefont {E.}~\bibnamefont {Tonni}},\ }\href
  {\doibase 10.1103/PhysRevLett.109.130502} {\bibfield  {journal} {\bibinfo
  {journal} {Phys. Rev. Lett.}\ }\textbf {\bibinfo {volume} {109}},\ \bibinfo
  {pages} {130502} (\bibinfo {year} {2012})}\BibitemShut {NoStop}%
\bibitem [{\citenamefont {Calabrese}\ \emph {et~al.}(2015)\citenamefont
  {Calabrese}, \citenamefont {Cardy},\ and\ \citenamefont
  {Tonni}}]{calabrese2015}%
  \BibitemOpen
  \bibfield  {author} {\bibinfo {author} {\bibfnamefont {P.}~\bibnamefont
  {Calabrese}}, \bibinfo {author} {\bibfnamefont {J.}~\bibnamefont {Cardy}}, \
  and\ \bibinfo {author} {\bibfnamefont {E.}~\bibnamefont {Tonni}},\ }\href
  {http://stacks.iop.org/1751-8121/48/i=1/a=015006} {\bibfield  {journal}
  {\bibinfo  {journal} {Journal of Physics A: Mathematical and Theoretical}\
  }\textbf {\bibinfo {volume} {48}},\ \bibinfo {pages} {015006} (\bibinfo
  {year} {2015})}\BibitemShut {NoStop}%
\bibitem [{\citenamefont {Wichterich}\ \emph {et~al.}(2009)\citenamefont
  {Wichterich}, \citenamefont {Molina-Vilaplana},\ and\ \citenamefont
  {Bose}}]{wichterich2009}%
  \BibitemOpen
  \bibfield  {author} {\bibinfo {author} {\bibfnamefont {H.}~\bibnamefont
  {Wichterich}}, \bibinfo {author} {\bibfnamefont {J.}~\bibnamefont
  {Molina-Vilaplana}}, \ and\ \bibinfo {author} {\bibfnamefont
  {S.}~\bibnamefont {Bose}},\ }\href {\doibase 10.1103/PhysRevA.80.010304}
  {\bibfield  {journal} {\bibinfo  {journal} {Phys. Rev. A}\ }\textbf {\bibinfo
  {volume} {80}},\ \bibinfo {pages} {010304} (\bibinfo {year}
  {2009})}\BibitemShut {NoStop}%
\bibitem [{\citenamefont {Ruggiero}\ \emph {et~al.}(2016)\citenamefont
  {Ruggiero}, \citenamefont {Alba},\ and\ \citenamefont
  {Calabrese}}]{ruggiero2016}%
  \BibitemOpen
  \bibfield  {author} {\bibinfo {author} {\bibfnamefont {P.}~\bibnamefont
  {Ruggiero}}, \bibinfo {author} {\bibfnamefont {V.}~\bibnamefont {Alba}}, \
  and\ \bibinfo {author} {\bibfnamefont {P.}~\bibnamefont {Calabrese}},\ }\href
  {\doibase 10.1103/PhysRevB.94.035152} {\bibfield  {journal} {\bibinfo
  {journal} {Phys. Rev. B}\ }\textbf {\bibinfo {volume} {94}},\ \bibinfo
  {pages} {035152} (\bibinfo {year} {2016})}\BibitemShut {NoStop}%
\bibitem [{\citenamefont {Lee}\ and\ \citenamefont {Vidal}(2013)}]{vidal2013}%
  \BibitemOpen
  \bibfield  {author} {\bibinfo {author} {\bibfnamefont {Y.~A.}\ \bibnamefont
  {Lee}}\ and\ \bibinfo {author} {\bibfnamefont {G.}~\bibnamefont {Vidal}},\
  }\href {\doibase 10.1103/PhysRevA.88.042318} {\bibfield  {journal} {\bibinfo
  {journal} {Phys. Rev. A}\ }\textbf {\bibinfo {volume} {88}},\ \bibinfo
  {pages} {042318} (\bibinfo {year} {2013})}\BibitemShut {NoStop}%
\bibitem [{\citenamefont {Castelnovo}(2013)}]{castelnovo2013}%
  \BibitemOpen
  \bibfield  {author} {\bibinfo {author} {\bibfnamefont {C.}~\bibnamefont
  {Castelnovo}},\ }\href {\doibase 10.1103/PhysRevA.88.042319} {\bibfield
  {journal} {\bibinfo  {journal} {Phys. Rev. A}\ }\textbf {\bibinfo {volume}
  {88}},\ \bibinfo {pages} {042319} (\bibinfo {year} {2013})}\BibitemShut
  {NoStop}%
\bibitem [{\citenamefont {Hart}\ and\ \citenamefont
  {Castelnovo}(2018)}]{castelnovo2018}%
  \BibitemOpen
  \bibfield  {author} {\bibinfo {author} {\bibfnamefont {O.}~\bibnamefont
  {Hart}}\ and\ \bibinfo {author} {\bibfnamefont {C.}~\bibnamefont
  {Castelnovo}},\ }\href {\doibase 10.1103/PhysRevB.97.144410} {\bibfield
  {journal} {\bibinfo  {journal} {Phys. Rev. B}\ }\textbf {\bibinfo {volume}
  {97}},\ \bibinfo {pages} {144410} (\bibinfo {year} {2018})}\BibitemShut
  {NoStop}%
\bibitem [{\citenamefont {Lu}\ and\ \citenamefont
  {Grover}(2019)}]{Lu2018_singularity}%
  \BibitemOpen
  \bibfield  {author} {\bibinfo {author} {\bibfnamefont {T.-C.}\ \bibnamefont
  {Lu}}\ and\ \bibinfo {author} {\bibfnamefont {T.}~\bibnamefont {Grover}},\
  }\href {\doibase 10.1103/PhysRevB.99.075157} {\bibfield  {journal} {\bibinfo
  {journal} {Phys. Rev. B}\ }\textbf {\bibinfo {volume} {99}},\ \bibinfo
  {pages} {075157} (\bibinfo {year} {2019})}\BibitemShut {NoStop}%
\bibitem [{\citenamefont {Gray}(2018)}]{gray2018fast}%
  \BibitemOpen
  \bibfield  {author} {\bibinfo {author} {\bibfnamefont {J.}~\bibnamefont
  {Gray}},\ }\href@noop {} {\bibfield  {journal} {\bibinfo  {journal} {arXiv
  preprint arXiv:1809.01685}\ } (\bibinfo {year} {2018})}\BibitemShut {NoStop}%
\bibitem [{\citenamefont {Hastings}\ \emph {et~al.}(2010)\citenamefont
  {Hastings}, \citenamefont {Gonz\'alez}, \citenamefont {Kallin},\ and\
  \citenamefont {Melko}}]{kallin2010}%
  \BibitemOpen
  \bibfield  {author} {\bibinfo {author} {\bibfnamefont {M.~B.}\ \bibnamefont
  {Hastings}}, \bibinfo {author} {\bibfnamefont {I.}~\bibnamefont
  {Gonz\'alez}}, \bibinfo {author} {\bibfnamefont {A.~B.}\ \bibnamefont
  {Kallin}}, \ and\ \bibinfo {author} {\bibfnamefont {R.~G.}\ \bibnamefont
  {Melko}},\ }\href {\doibase 10.1103/PhysRevLett.104.157201} {\bibfield
  {journal} {\bibinfo  {journal} {Phys. Rev. Lett.}\ }\textbf {\bibinfo
  {volume} {104}},\ \bibinfo {pages} {157201} (\bibinfo {year}
  {2010})}\BibitemShut {NoStop}%
\bibitem [{\citenamefont {Calabrese}\ \emph {et~al.}(2013)\citenamefont
  {Calabrese}, \citenamefont {Cardy},\ and\ \citenamefont
  {Tonni}}]{calabrese2013entanglement}%
  \BibitemOpen
  \bibfield  {author} {\bibinfo {author} {\bibfnamefont {P.}~\bibnamefont
  {Calabrese}}, \bibinfo {author} {\bibfnamefont {J.}~\bibnamefont {Cardy}}, \
  and\ \bibinfo {author} {\bibfnamefont {E.}~\bibnamefont {Tonni}},\
  }\href@noop {} {\bibfield  {journal} {\bibinfo  {journal} {Journal of
  Statistical Mechanics: Theory and Experiment}\ }\textbf {\bibinfo {volume}
  {2013}},\ \bibinfo {pages} {P02008} (\bibinfo {year} {2013})}\BibitemShut
  {NoStop}%
\bibitem [{\citenamefont {Alba}(2013)}]{alba2013}%
  \BibitemOpen
  \bibfield  {author} {\bibinfo {author} {\bibfnamefont {V.}~\bibnamefont
  {Alba}},\ }\href {\doibase 10.1088/1742-5468/2013/05/p05013} {\bibfield
  {journal} {\bibinfo  {journal} {Journal of Statistical Mechanics: Theory and
  Experiment}\ }\textbf {\bibinfo {volume} {2013}},\ \bibinfo {pages} {P05013}
  (\bibinfo {year} {2013})}\BibitemShut {NoStop}%
\bibitem [{\citenamefont {Chung}\ \emph {et~al.}(2014)\citenamefont {Chung},
  \citenamefont {Alba}, \citenamefont {Bonnes}, \citenamefont {Chen},\ and\
  \citenamefont {L\"auchli}}]{Chiamin:2014repqmc}%
  \BibitemOpen
  \bibfield  {author} {\bibinfo {author} {\bibfnamefont {C.-M.}\ \bibnamefont
  {Chung}}, \bibinfo {author} {\bibfnamefont {V.}~\bibnamefont {Alba}},
  \bibinfo {author} {\bibfnamefont {L.}~\bibnamefont {Bonnes}}, \bibinfo
  {author} {\bibfnamefont {P.}~\bibnamefont {Chen}}, \ and\ \bibinfo {author}
  {\bibfnamefont {A.~M.}\ \bibnamefont {L\"auchli}},\ }\href {\doibase
  10.1103/PhysRevB.90.064401} {\bibfield  {journal} {\bibinfo  {journal} {Phys.
  Rev. B}\ }\textbf {\bibinfo {volume} {90}},\ \bibinfo {pages} {064401}
  (\bibinfo {year} {2014})}\BibitemShut {NoStop}%
\bibitem [{sup()}]{supplement}%
  \BibitemOpen
  \href@noop {} {\ }\bibinfo {note} {See supplemental material}\BibitemShut
  {NoStop}%
\bibitem [{\citenamefont {Kitaev}\ and\ \citenamefont
  {Preskill}(2006)}]{Kitaev06_1}%
  \BibitemOpen
  \bibfield  {author} {\bibinfo {author} {\bibfnamefont {A.}~\bibnamefont
  {Kitaev}}\ and\ \bibinfo {author} {\bibfnamefont {J.}~\bibnamefont
  {Preskill}},\ }\href {\doibase 10.1103/PhysRevLett.96.110404} {\bibfield
  {journal} {\bibinfo  {journal} {Phys. Rev. Lett.}\ }\textbf {\bibinfo
  {volume} {96}},\ \bibinfo {pages} {110404} (\bibinfo {year}
  {2006})}\BibitemShut {NoStop}%
\bibitem [{\citenamefont {Levin}\ and\ \citenamefont
  {Wen}(2006)}]{levin2006detecting}%
  \BibitemOpen
  \bibfield  {author} {\bibinfo {author} {\bibfnamefont {M.}~\bibnamefont
  {Levin}}\ and\ \bibinfo {author} {\bibfnamefont {X.-G.}\ \bibnamefont
  {Wen}},\ }\href {\doibase 10.1103/PhysRevLett.96.110405} {\bibfield
  {journal} {\bibinfo  {journal} {Phys. Rev. Lett.}\ }\textbf {\bibinfo
  {volume} {96}},\ \bibinfo {pages} {110405} (\bibinfo {year}
  {2006})}\BibitemShut {NoStop}%
\bibitem [{\citenamefont {Hesselmann}\ and\ \citenamefont
  {Wessel}(2016)}]{Wessel:2016tfim}%
  \BibitemOpen
  \bibfield  {author} {\bibinfo {author} {\bibfnamefont {S.}~\bibnamefont
  {Hesselmann}}\ and\ \bibinfo {author} {\bibfnamefont {S.}~\bibnamefont
  {Wessel}},\ }\href {\doibase 10.1103/PhysRevB.93.155157} {\bibfield
  {journal} {\bibinfo  {journal} {Phys. Rev. B}\ }\textbf {\bibinfo {volume}
  {93}},\ \bibinfo {pages} {155157} (\bibinfo {year} {2016})}\BibitemShut
  {NoStop}%
\bibitem [{\citenamefont {Ferdinand}\ and\ \citenamefont
  {Fisher}(1969)}]{fisher1969}%
  \BibitemOpen
  \bibfield  {author} {\bibinfo {author} {\bibfnamefont {A.~E.}\ \bibnamefont
  {Ferdinand}}\ and\ \bibinfo {author} {\bibfnamefont {M.~E.}\ \bibnamefont
  {Fisher}},\ }\href {\doibase 10.1103/PhysRev.185.832} {\bibfield  {journal}
  {\bibinfo  {journal} {Phys. Rev.}\ }\textbf {\bibinfo {volume} {185}},\
  \bibinfo {pages} {832} (\bibinfo {year} {1969})}\BibitemShut {NoStop}%
\bibitem [{\citenamefont {Sherman}\ \emph {et~al.}(2016)\citenamefont
  {Sherman}, \citenamefont {Devakul}, \citenamefont {Hastings},\ and\
  \citenamefont {Singh}}]{sherman2016}%
  \BibitemOpen
  \bibfield  {author} {\bibinfo {author} {\bibfnamefont {N.~E.}\ \bibnamefont
  {Sherman}}, \bibinfo {author} {\bibfnamefont {T.}~\bibnamefont {Devakul}},
  \bibinfo {author} {\bibfnamefont {M.~B.}\ \bibnamefont {Hastings}}, \ and\
  \bibinfo {author} {\bibfnamefont {R.~R.~P.}\ \bibnamefont {Singh}},\ }\href
  {\doibase 10.1103/PhysRevE.93.022128} {\bibfield  {journal} {\bibinfo
  {journal} {Phys. Rev. E}\ }\textbf {\bibinfo {volume} {93}},\ \bibinfo
  {pages} {022128} (\bibinfo {year} {2016})}\BibitemShut {NoStop}%
\bibitem [{\citenamefont {Stoudenmire}\ and\ \citenamefont
  {White}(2010)}]{Stoudenmire_2010}%
  \BibitemOpen
  \bibfield  {author} {\bibinfo {author} {\bibfnamefont {E.~M.}\ \bibnamefont
  {Stoudenmire}}\ and\ \bibinfo {author} {\bibfnamefont {S.~R.}\ \bibnamefont
  {White}},\ }\href {\doibase 10.1088/1367-2630/12/5/055026} {\bibfield
  {journal} {\bibinfo  {journal} {New Journal of Physics}\ }\textbf {\bibinfo
  {volume} {12}},\ \bibinfo {pages} {055026} (\bibinfo {year}
  {2010})}\BibitemShut {NoStop}%
\bibitem [{\citenamefont {Dennis}\ \emph {et~al.}(2002)\citenamefont {Dennis},
  \citenamefont {Kitaev}, \citenamefont {Landahl},\ and\ \citenamefont
  {Preskill}}]{dennis2002}%
  \BibitemOpen
  \bibfield  {author} {\bibinfo {author} {\bibfnamefont {E.}~\bibnamefont
  {Dennis}}, \bibinfo {author} {\bibfnamefont {A.}~\bibnamefont {Kitaev}},
  \bibinfo {author} {\bibfnamefont {A.}~\bibnamefont {Landahl}}, \ and\
  \bibinfo {author} {\bibfnamefont {J.}~\bibnamefont {Preskill}},\ }\href
  {\doibase 10.1063/1.1499754} {\bibfield  {journal} {\bibinfo  {journal}
  {Journal of Mathematical Physics}\ }\textbf {\bibinfo {volume} {43}},\
  \bibinfo {pages} {4452} (\bibinfo {year} {2002})},\ \Eprint
  {http://arxiv.org/abs/https://doi.org/10.1063/1.1499754}
  {https://doi.org/10.1063/1.1499754} \BibitemShut {NoStop}%
\bibitem [{\citenamefont {Serafini}(2017)}]{serafini2017quantum}%
  \BibitemOpen
  \bibfield  {author} {\bibinfo {author} {\bibfnamefont {A.}~\bibnamefont
  {Serafini}},\ }\href@noop {} {\emph {\bibinfo {title} {Quantum Continuous
  Variables: A Primer of Theoretical Methods}}}\ (\bibinfo  {publisher} {CRC
  Press},\ \bibinfo {year} {2017})\BibitemShut {NoStop}%
\bibitem [{\citenamefont {Iba}(2001)}]{Iba:EEMC}%
  \BibitemOpen
  \bibfield  {author} {\bibinfo {author} {\bibfnamefont {Y.}~\bibnamefont
  {Iba}},\ }\href {\doibase 10.1142/S0129183101001912} {\bibfield  {journal}
  {\bibinfo  {journal} {International Journal of Modern Physics C}\ }\textbf
  {\bibinfo {volume} {12}},\ \bibinfo {pages} {623} (\bibinfo {year} {2001})},\
  \Eprint {http://arxiv.org/abs/https://doi.org/10.1142/S0129183101001912}
  {https://doi.org/10.1142/S0129183101001912} \BibitemShut {NoStop}%
\end{thebibliography}%

\section{Supplemental Material}

\section{Renyi negativity in the quantum spherical model }\label{appendix:spherical}
The quantum spherical model is described by the Hamiltonian $H=\frac{1}{2} g \sum_{i=1}^N p_i^2 -\frac{1}{2N} \sum_{i,j=1}^N x_ix_j $, with $[x_i,p_j] = i \delta_{ij}$, and the spherical constraint $\delta\left( \frac{1}{N} \sum_{i=1}^N x_i^2 -\frac{1}{4}  \right)$. Employing a standard path integral representation for the partition function at the inverse temperature $\beta$, the constraint induces a term $\mu \left[  \sum_{i=1}^N x_i^2-\frac{N}{4} \right]$ in the action, where the Lagrange multiplier $\mu$ needs to be integrated over. As $N\to \infty$, one can perform a saddle point approximation to neglect the fluctuation of $\mu$, resulting in a Gaussian theory with the effective Hamiltonian
$H=\frac{1}{2} g \sum_{i=1}^N p_i^2 -\frac{1}{2N} \sum_{i,j=1}^N x_ix_j  + \mu \left[ \sum_{i=1}^N x_i^2 -\frac{N}{4} \right] $. $\mu$ is the saddle point solution, chosen so that $\left< \sum_{i=1}^N x_i^2 \right>_{\beta} =\frac{N}{4}$, where the expectation value is taken with respect to the Gibbs state $\rho\sim e^{-\beta H}$ at the inverse temperature $\beta$.

Here we divide the system into the two subsystems A and B of equal size, and study the Renyi negativity $R_n$, defined as
\begin{equation}
R_n=   - \log \left\{      \frac{  \tr\left[ \left(   \rho^{T_B}  \right)^n \right] }{    \tr \rho^n}  \right\}   =-\log \left\{      \frac{  \tr \left[  \left(  e^{-\beta H }  \right)^{T_B} \right]^n }{\tr e^{-n\beta H}}  \right\}.
\end{equation}
Below we show that while negativity is singular at $\beta_c$, the critical inverse temperature corresponding to the state $\sim e^{-\beta H}$\cite{Lu2018_singularity}, $R_n$ exhibits a singularity at the inverse temperature $\beta_c/n$. First we note that $R_n$ involves the partition function at inverse temperature $n\beta$. That implies in the calculation of $R_n$, $\mu$ is chosen so that $\left< \sum_{i=1}^N x_i^2 \right>_{n\beta} =\frac{N}{4}$ where the expectation value is taken with respect to the Gibbs state at $n\beta$, i.e. $\sim e^{-n\beta H}$. 

\begin{figure}
	\centering
	\includegraphics[width=\linewidth]{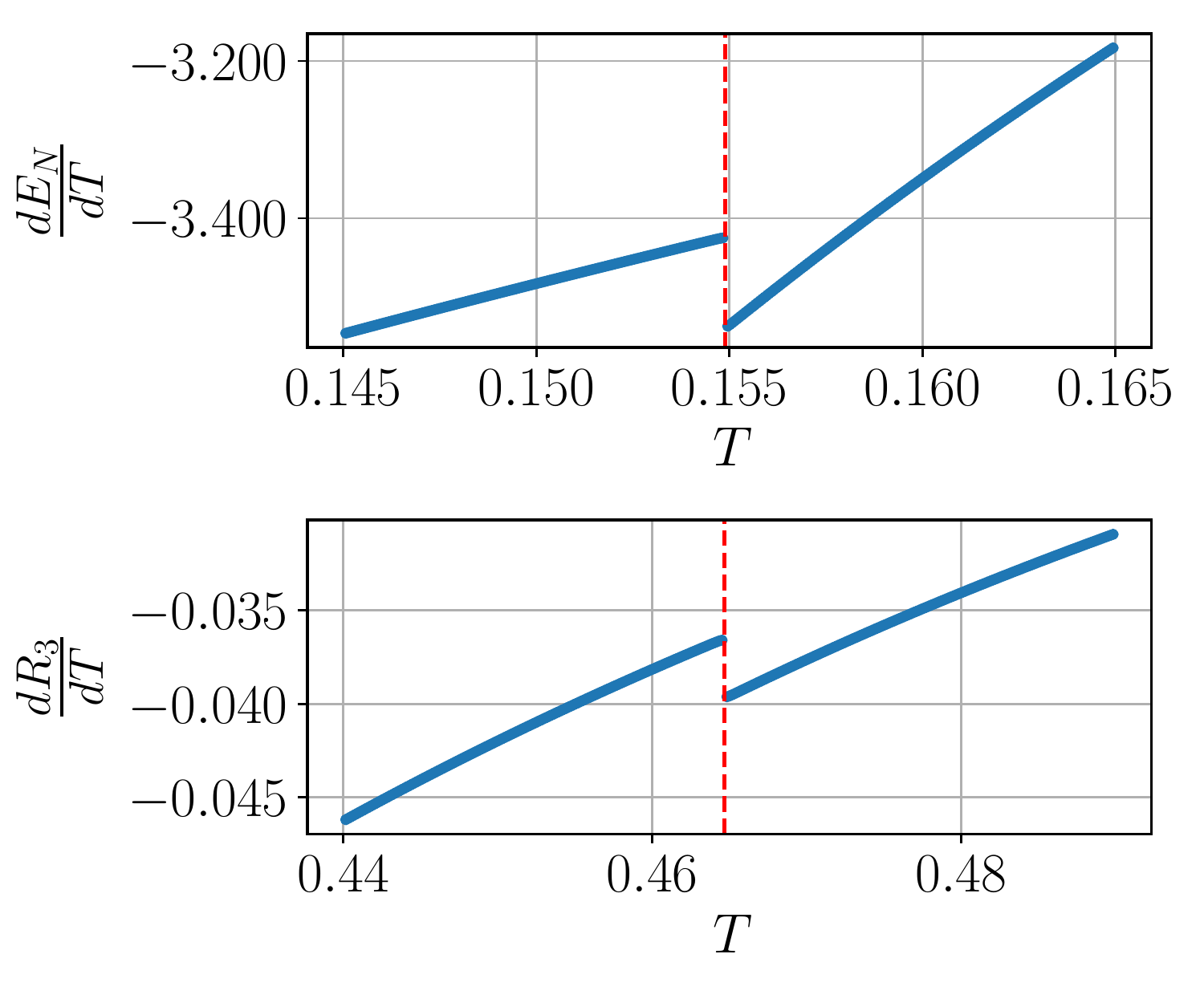}
	\caption{The upper panel and the lower panel show the temperature derivative of  negativity $E_N$ and the third Renyi negativity $R_3$ respectively across the critical point in the quantum spherical model. The vertical dashed line in the upper (lower) panel  indicates the temperature $T_c$ ($3T_c$), where $T_c$ is critical temperature.}
	\label{fig:spherical}
\end{figure}

Following the calculation in Ref.~\cite{Lu2018_singularity}, we find for $2\sqrt{g} \coth(\frac{1}{2}n\beta \sqrt{g})>1$, the system is in the disordered phase, with $\mu$ determined from $\sqrt{\frac{g}{2\mu}} \coth(\frac{1}{2}n\beta \sqrt{2g\mu})=\frac{1}{2}$
while the condition $2\sqrt{g} \coth(\frac{1}{2}n\beta \sqrt{g})<1$ gives the ordered phase, and $\mu$ is pinned to $\frac{1}{2}$. Having determined $\mu$, now we can calculate $R_n$ using the covariance matrix technique since both $\rho$ and $\rho^{T_B}$ are Gaussian states. Using the result from Ref.~\cite{serafini2017quantum}, one finds
\begin{equation}\label{spherical:renyi}
R_n =-\sum_{i=1}^N  \log \left[   \frac{  (\nu_i+1)^n - (\nu_i-1)^n  }{  (  \tilde{\nu}_i+1   )^n -  (    \tilde{\nu}_i-1)^n      }   \right],
\end{equation} 
where $\{  \nu_i\}$ and $\{\tilde{\nu}_i  \}$ are the symplectic spectra of the covariance matrices from $\rho$ and $\rho^{T_B}$ respectively. A calculation similar to Ref.~\cite{Lu2018_singularity} gives the symplectic spectra

\begin{equation}
\nu_i= \begin{cases}
\coth(  \frac{1}{2} \beta \sqrt{2\mu g}  )  ~ \text{for } ~ i=1,\cdots, N-1\\
\coth(  \frac{1}{2} \beta \sqrt{(2\mu -1)g}  )   ~ \text{for } ~ i=N
\end{cases}
\end{equation}

\begin{equation}
\begin{split}
&\tilde{\nu}_i=\\
& \begin{cases}
\coth\left( \frac{1}{2}\beta \sqrt{2\mu g }   \right) ~ \text{for} ~  i=1,2, \cdots N-2\\
\left[\sqrt{\frac{2\mu}{2\mu-1}}  \coth\left(\frac{1}{2} \beta \sqrt{ (2\mu-1)g}  \right)  \coth\left( \frac{1}{2}\beta \sqrt{2\mu g }   \right) \right]^{\frac{1}{2}}  ~ \text{for}~  i=N-1  \\
\left[\sqrt{\frac{2\mu-1}{2\mu}}  \coth\left(\frac{1}{2} \beta \sqrt{ (2\mu-1)g}  \right)  \coth\left( \frac{1}{2}\beta \sqrt{2\mu g }   \right) \right]^{\frac{1}{2}}      ~ \text{for} ~ i=N 
\end{cases}
\end{split}
\end{equation}
Plugging $\{\nu_i\}$ and $\{\tilde{\nu}_i\}$  in Eq.~\eqref{spherical:renyi} and choosing $n=3$, one finds 
\begin{equation}\label{appendix:renyi_result}
\boxed{
	R_3= -   \sum_{i=1}^2\log \left[    \frac{3\lambda_i+1}{3\tilde{\lambda}_i+1   }  \right]}
\end{equation}
where 

\begin{equation}
\boxed{
	\begin{split}
	&\lambda_1=\coth^2\left(  \frac{1}{2} \beta \sqrt{2\mu g}  \right) \\
	&\lambda_2=\coth^2\left(  \frac{1}{2} \beta \sqrt{(2\mu -1)g}  \right) \\
	& \tilde{\lambda}_1 =  \sqrt{\frac{2\mu}{2\mu-1}}  \coth\left(\frac{1}{2} \beta \sqrt{ (2\mu-1)g}  \right)  \coth\left( \frac{1}{2}\beta \sqrt{2\mu g }   \right) \\
	&  \tilde{\lambda}_2  =   \sqrt{\frac{2\mu-1}{2\mu}}  \coth\left(\frac{1}{2} \beta \sqrt{ (2\mu-1)g}  \right)  \coth\left( \frac{1}{2}\beta \sqrt{2\mu g }   \right)      \\
	\end{split}}
\end{equation} 
Since $\mu$ is singular at the inverse temperature $\beta_c/3$, the Renyi negativity is singular as well (see Fig.~\ref{fig:spherical}).

\section{Renyi negativity at the Gaussian fixed point}\label{App:gaussian}
To understand the universal quantum correlation across a finite temperature phase transition, Ref.~\cite{Lu2019_gaussian} studied a two dimensional model, which can be regarded as the mean-field approximation of the two dimensional transverse field Ising model while taking into account Gaussian fluctuations. Defining the canonically conjugate pair $(\phi_{\vec{r}}, \pi_{\vec{r}})$ at lattice site $\vec{r}$, the model Hamiltonian reads
\begin{equation}\label{eq:H}
H=\frac{1}{2} \sum_{\vec{r}} \left( \pi_{\vec{r}}^2 +m^2 \phi_{\vec{r}}^2 \right) +  \frac{1}{2}  \sum_{\expval{\vec{r},\vec{r}'}}K \left( \phi_{\vec{r}} -\phi_{\vec{r}'}   \right) ^2,
\end{equation}
where the physical mass $m$ obeys
\begin{equation}
m(T)=\begin{cases}
\sqrt{T-T_{n,c}} \quad \quad ~ \text{for} \quad T>T_{n,c}\\
\sqrt{2(T_{n,c}-T)}  \quad \text{for} \quad T<T_{n,c}.
\end{cases}
\end{equation}
$T_{n,c}=nT_c$ is the critical temperature corresponding to the state $\rho \sim e^{-n\beta H}$, where correlation length diverges due to the vanishing physical mass. Note that Ref.~\cite{Lu2019_gaussian} considers $m$ as a function of another tuning parameter $g$ (can be thought of as the transverse field in the quantum Ising model), but it leads to the same qualitative features as turning $T$. 

\begin{figure}
	\centering
	\includegraphics[width=\linewidth]{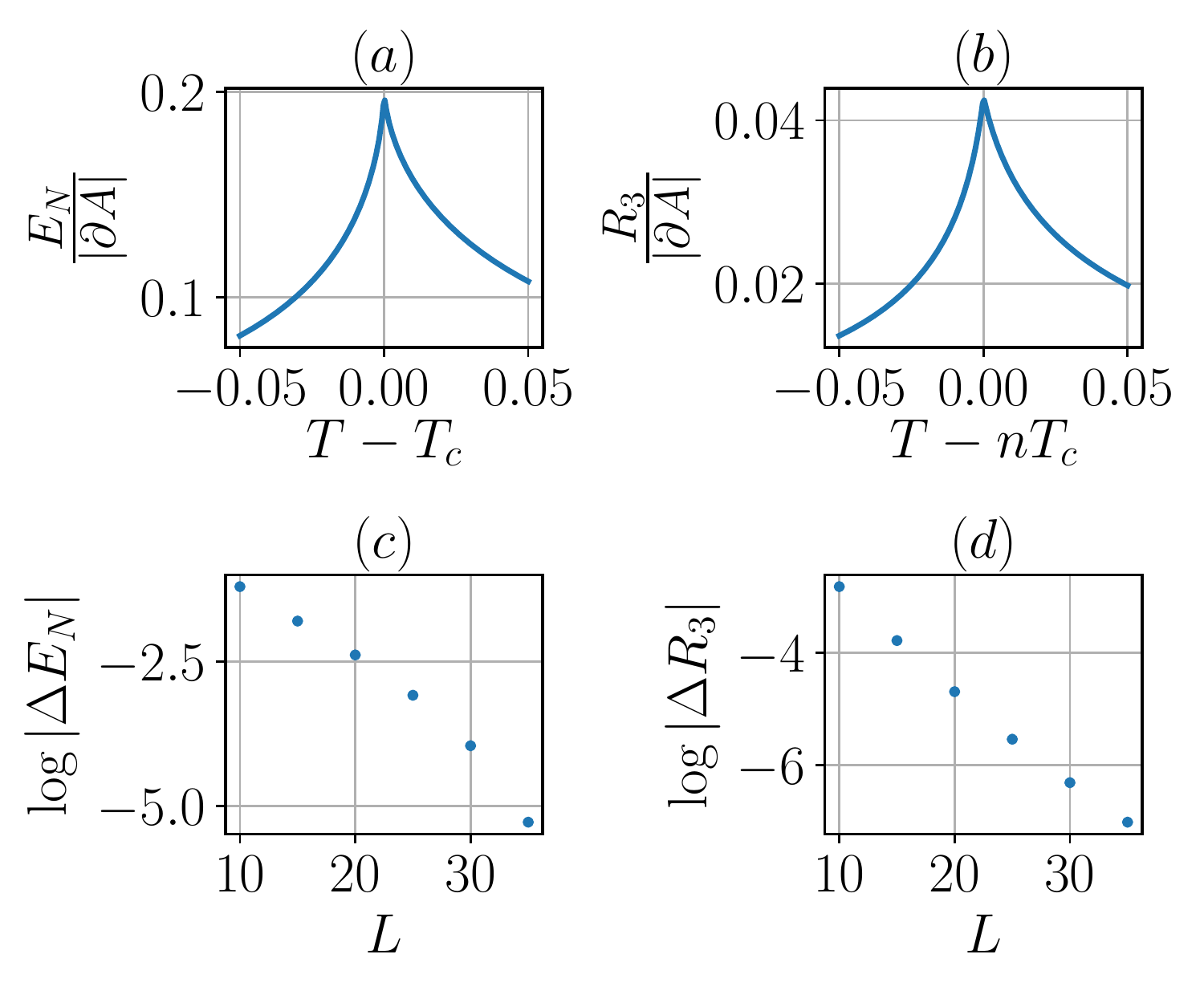}
	\caption{(a) and (b) show the singularity of the area law coefficients in negativity $E_N$ and the third Renyi negativity $R_3$ at the temperature $T_c$ and $3T_c$ respectively. $|\partial A|$ is the boundary length of a subregion $A$. (c) and (d) show the finite-size scaling of $E_N$ and $R_3$ at $T_c$ and $3T_c$.}
	\label{fig:gaussian}
\end{figure}

By choosing a square region of size $L_A=\frac{4}{5}L$, we find the area-law coefficients of negativity $E_N$ and Renyi negativity $R_3$ both exhibit a cusp singularity at the corresponding critical point temperature:$T_c$ and $3T_c$ as shown in Fig.~\ref{fig:gaussian}(a), (b). Next, to justify the use of $R_3$ for capturing the universal component of quantum correlations, we employ the Kitaev-Preskill subtraction scheme \cite{Kitaev06_1}, where the subregions $A$, $B$, and $C$ considered in this scheme are squares of size $2/5L, 2/5L, 4/5L$ respectively. We find that even right at the critical temperature where the physical correlation length diverges, the subtracted Renyi negativity $\Delta R_3$ decays exponentially with the system size, similar to the behavior of subtracted negativity $\Delta E_N$ (Fig.~\ref{fig:gaussian}(c), (d)).

\section{Critical temperature of the 2D TFIM} \label{App:locate_Tc}
\begin{figure}[h!]
	\includegraphics[width=\linewidth]{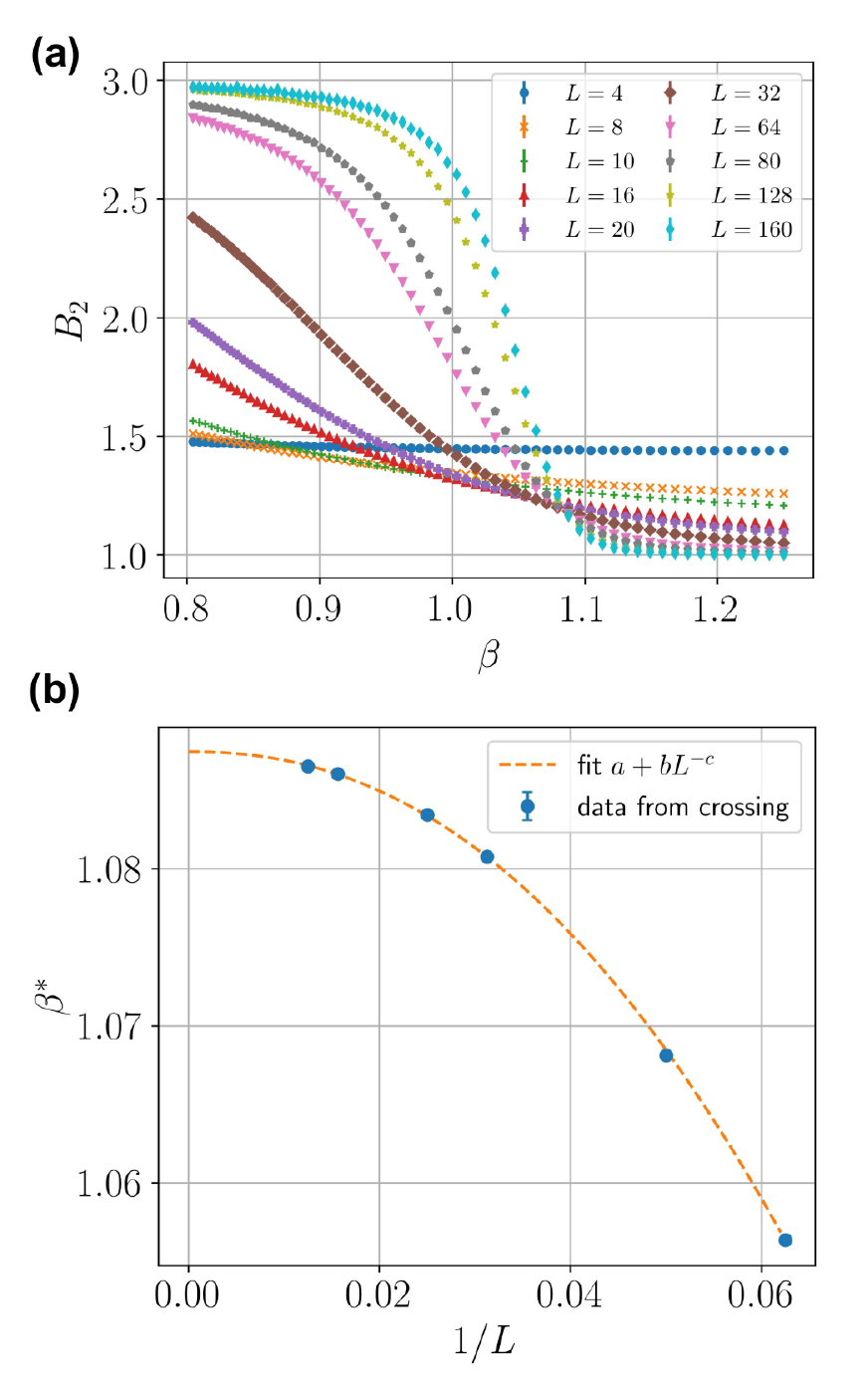}
	\caption{ \textbf{(a)} The Binder-ratio calculated using the standard SSE with $h_x = 2.75$. \textbf{(b)} The crossing inverse-temperature obtained from system sizes $L$ and $2L$ as a function of inverse system size. By fitting with a power-law $y = a + bL^{-c}$, the best fit occurs at $a = \beta_c = 1.0874(1)$, $b = -14(1)$ and $c = 2.21(3)$.}
	\label{Fig:B2} 
\end{figure}

Here we provide  details of the QMC simulation of the 2D  transverse field Ising model (TFIM). To locate the critical temperature in thermodynamic limit, we measured Binder ratio $B_2$  defined as
\begin{equation}
B_2 = \frac{\left<M_z^4\right>}{\left<M_z^2\right>^2}.
\end{equation}
Fig.~\ref{Fig:B2}(a) shows the Binder ratio $B_2$ with various system sizes at transverse field $H_x = 2.75$.  The critical temperature can be extracted with the finite-size scaling using the crossing analysis discussed below, which yields $\beta_c = 1.0874(1)$

\noindent\textit{Crossing analysis:}
\quad  In the following, we describe the procedure to locate the critical temperature $T_c$ using finite-size scaling. Here, we consider a dimension less quantity $R$ (such as Binder ratio $B_2$). The standard finite-size scaling function with system size $L$ , defines as,
\begin{equation}
R(b,L) = f(bL^{\frac{1}{\nu}})(1+\alpha L^{-\omega})
\end{equation}
where $b \equiv \beta - \beta_c$ is the reduced inverse temperature, $\nu$ is the correlation exponent, and $\omega$ is the correction to scaling exponent. 

Near the transition where $b \ll 1$, one can Taylor-expand the $R(b)$ as
\begin{equation}
R(b,L) \approx \left[  R_{c}^{\infty} + abL^{\frac{1}{\nu}}  \right] (1+\alpha L^{-\omega}).
\end{equation} 
Considering two curves with system size $L$ and $nL$, one has
\begin{align}
R(b,L) &= \left[  R_{c}^{\infty} + a b L^{\frac{1}{\nu}}  \right] (1+\alpha L^{-\omega}), \label{eq:R}\\
R(b,nL) &= \left[  R_{c}^{\infty} + a b L^{\frac{1}{\nu}}n^{\frac{1}{\nu}}  \right] (1+\alpha L^{-\omega}n^{-\omega}). \label{eq:nR} 		
\end{align} 
At the crossing point of the two curves, $R(b,L) = R(b,nL)$, from which one finds the crossing inverse temperature $b$ as
\begin{equation}
b = \frac{\alpha R_c^{\infty} L^{-\omega}(1-n^{-\omega})}{aL^{\frac{1}{\nu}}\left[(n^{\frac{1}{\nu}} - 1) + \alpha(n^{\frac{1}{\nu} - \omega}) L^{-\omega}\right]} 
\approx 
\frac{\alpha R_c^{\infty} L^{-\omega}(1-n^{-\omega})}{aL^{\frac{1}{\nu}}(n^{\frac{1}{\nu}} - 1)}. 
\end{equation} 
In the last step, we ignore the sub-leading term in the denominator that will eventually go to zero as $L\rightarrow \infty$. Finally, the crossing inverse temperature $\beta^*$ with pair $L$ and $nL$ can be derived as 
\begin{equation}
\beta^*(L) = \beta_c + c(n)L^{-\omega-\frac{1}{\nu}}.\label{eq:crxT}.
\end{equation}
Inserting Eq.~\eqref{eq:crxT} into Eq.~\eqref{eq:R}, one can get the crossing quantity $R^*$ as
\begin{equation}
R^{*}(L) = R_{c} + d(n)L^{-\omega}.
\end{equation}
Where the above $c(n)$ and $d(n)$ are $n$ dependent coefficients.

\section{Computation of  $\gamma$  with expanded ensemble and re-weighting method} \label{App:Reweight}

In the following, we show how to measure the $\gamma$ in a single QMC simulation without numerical integration or subtraction of the estimated data from multiple QMC simulations. 

We start with Eq.~\eqref{eq:Zratio}, where the extended partition function $Z[A,\beta,3]$ and $Z[3\beta]$ are simulated using the replica trick by SSE. 
For $Z[A,\beta,3]$, the boundary condition in the imaginary-time direction is modified to represent the partial transpose as shown in Fig.~\ref{Fig:TopoZZ3}. To estimate $\gamma$, we use Levin-Wen subtraction scheme (Ref.~\cite{levin2006detecting}, Eq.~\ref{eq:red}). Note that after the subtraction, all factors of $Z[3\beta]$  cancel out, leaving only the $Z_{S_i} = Z[S_i,\beta,3]$ and, 
\begin{equation}
\gamma = \frac{1}{2}\log\left(\frac{Z_{S_2}^2}{Z_{S_1}Z_{S_3}} \right),
\end{equation} 
where $S_i$ are the sub-regions defined in the main text.

The measurement of $\gamma$ can be directly calculated in QMC by an expanded ensemble method ~\cite{Iba:EEMC}. The idea is to combine partition functions of different subregions $S_i$ into a single simulation, by defining
\begin{equation}
Z_\text{tot} = \sum_i Z_{S_i},
\end{equation} 
where $Z_\text{tot}$ is the sum of the partition functions and $S_i$ is the subregion as shown in the inset of Fig.~\ref{Fig:LWRes}.

\begin{figure}[h!]
	\includegraphics[width=\linewidth]{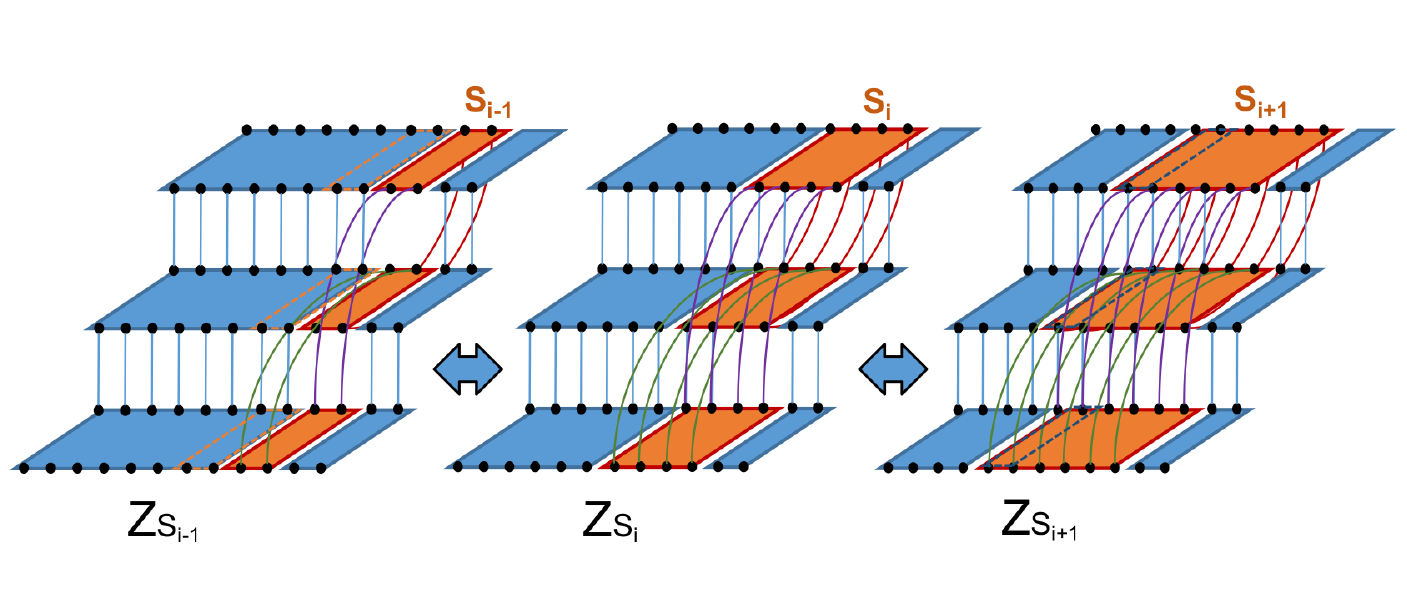}
	\caption{Illustration of the change of imaginary-time boundary conditions. The orange parts indicate the sub-region $S_j$, and the regions with dash-line indicate the $S_i \oplus S_{i\pm1}$. Starting with $Z_{S_i}$, one can propose the update with equal probability to either $Z_{S_{i-1}}$ (left) by changing the imaginary time boundary on light dash-line region; or to $Z_{S_{i+1}}$(right) by changing the imaginary time boundary on dark dash-line region.} 
	\label{Fig:Upd} 
\end{figure}

The simulation starts in a certain partition function $Z_{S_i}$. In each Monte-Carlo step, we first perform a standard SSE update to update the spin configuration. We then propose another update to switch from partition $Z_{S_i}$ to $Z_{S_{i+1}}$ or $Z_{S_{i-1}}$ with equal probability. The update essentially  modifies the imaginary-time boundary conditions in the update region $S_{i} \oplus S_{i \pm 1}$ as shown in Fig.\ref{Fig:Upd}. The switch can only be accepted if the current configuration remains consistent with the new boundary conditions. For example, if one choose to start with $Z_{S_2}$, the switch can update from $Z_{S_2}$ to $Z_{S_1}$ by proposing the change in imaginary-time boundary on region $\Xi_2$; or update from $Z_{S_2}$ to $Z_{S_3}$ by proposing the change on region $\Xi_3$. 

However, one can easily see that problem arises when we increase the system size. The acceptance rate for the switching between  two regions $S_i$ and $S_j$ is proportional to the ratio of the partition functions $Z_{S_j}/Z_{S_i}$. Since the Renyi negativity obeys an area-law scaling, one expects $\Delta R_3 = -\ln \frac{Z_{S_j}}{Z_{S_i}} = c\Delta l$, where $c$ is a constant and $\Delta l$ is the difference in the boundary length between sub-regions $S_i$ and $S_j$. 

To solve the problem, for a larger system size we perform two modifications in our simulations. First, we add intermediate ensembles to help tunneling between two desired ensembles $Z_{S_i}$ and $Z_{S_j}$. Instead of attempting an update of the whole region $\Xi_i$ at once, we introduce additional ensembles by further partitioning $\Xi_i$ into smaller parts.    Second, we implement a re-weighting scheme to optimize the sampling efficiency. The idea is very similar to the spirit of simulated tempering. At the beginning of the simulation, we first iteratively search  for the weight $g$ such that the ratio between the weighted partition functions is roughly one,
\begin{equation}
\frac{g_j Z_{S_j}}{g_i Z_{S_i}} \sim 1.
\end{equation} 

We then fix ${g_i}$ and perform the simulation with the modified partition function 
\begin{equation}
Z^{'}_\text{tot} = \sum_i g_i Z_{S_i}.
\end{equation}
The partition function ratio is estimated by counting how many times the simulation is in the desired set-up $S_i$, and the physical estimators can be calculated by re-weighting the ratio in post data processing as  
\begin{equation}
\frac{Z_{S_j}}{Z_{S_i}} = \frac{g_i\left\langle N_j\right\rangle}{g_j\left\langle N_i\right\rangle}.
\end{equation}

\end{document}